\documentclass[sigconf,authorversion,nonacm]{acmart}

\setcopyright{none}

\usepackage{todonotes}
\usepackage{tabularx}
\usepackage{import}
\usetikzlibrary{quotes,arrows.meta}
\usetikzlibrary{positioning}

\def\edgecolor{rgb:blue,4;red,1;green,4;black,3}
\newcommand{\midarrow}{\tikz \draw[-Stealth,line width =0.8mm,draw=\edgecolor] (-0.3,0) -- ++(0.3,0);}

\usepackage{Ball}
\usepackage{Box}
\usepackage{RightBandedBox}

\usetikzlibrary{positioning}
\usetikzlibrary{3d} 

\def\ConvColor{rgb:yellow,5;red,2.5;white,5}

\def\PoolColor{rgb:red,1;black,0.3}
\def\UnpoolColor{rgb:blue,2;green,1;black,0.3}
\def\FcColor{rgb:blue,5;red,2.5;white,5}

\def\LstmColor{rgb:blue,5;green,15}

\usepackage{array}
\newcolumntype{L}[1]{>{\raggedright\let\newline\\\arraybackslash\hspace{0pt}}m{#1}}
\newcolumntype{C}[1]{>{\centering\let\newline\\\arraybackslash\hspace{0pt}}m{#1}}
\newcolumntype{R}[1]{>{\raggedleft\let\newline\\\arraybackslash\hspace{0pt}}m{#1}}

\newcommand{\sysname}{TouchFusion}
\begin{document}

\title{\sysname{}: Multimodal Wristband Sensing for Ubiquitous Touch Interactions}
\author{Eric Whitmire}
\email{ewhitmire@meta.com}
\affiliation{
  \institution{Meta}
  \country{}
}

\author{Evan Strasnick}
\affiliation{
  \institution{Meta}
  \country{}
}
\author{Roger Boldu}
\affiliation{
  \institution{Meta}
  \country{}
}
\author{Raj Sodhi}
\affiliation{
  \institution{Meta}
  \country{}
}
\author{Nathan Godwin}
\affiliation{
  \institution{Meta}
  \country{}
}
\author{Shiu Ng}
\affiliation{
  \institution{Meta}
  \country{}
}
\author{Andre Levi}
\affiliation{
  \institution{Meta}
  \country{}
}
\author{Amy Karlson}
\affiliation{
  \institution{Meta}
  \country{}
}
\author{Ran Tan}
\affiliation{
  \institution{Meta}
  \country{}
}
\author{Josef Faller}
\affiliation{
  \institution{Meta}
  \country{}
}
\author{Emrah Adamey}
\affiliation{
  \institution{Meta}
  \country{}
}
\author{Hanchuan Li}
\affiliation{
  \institution{Meta}
  \country{}
}
\author{Wolf Kienzle}
\affiliation{
  \institution{Meta}
  \country{}
}
\author{Hrvoje Benko}
\affiliation{
  \institution{Meta}
  \country{}
}

\renewcommand{\shortauthors}{Whitmire et al.}

\begin{abstract}
\sysname{} is a wristband that enables touch interactions on nearby surfaces without any additional instrumentation or computer vision. \sysname{} combines surface electromyography (sEMG), bioimpedance, inertial, and optical sensing to capture multiple facets of hand activity during touch interactions. Through a combination of early and late fusion, \sysname{} enables stateful touch detection on both environmental and body surfaces, simple surface gestures, and tracking functionality for contextually adaptive interfaces as well as basic trackpad-like interactions. We validate our approach on a dataset of 100 participants, significantly exceeding the population size of typical wearable sensing studies to capture a wider variance of wrist anatomies, skin conductivities, and behavioral patterns. We show that \sysname{} can enable several common touch interaction tasks. Using \sysname{}, a wearer can summon a trackpad on any surface, control contextually adaptive interfaces based on where they tap, or use their palm as an always-available touch surface. When paired with smart glasses or augmented reality devices, \sysname{} enables a ubiquitous, contextually adaptive interaction model.
\end{abstract}




\begin{teaserfigure}
  \includegraphics[width=\textwidth]{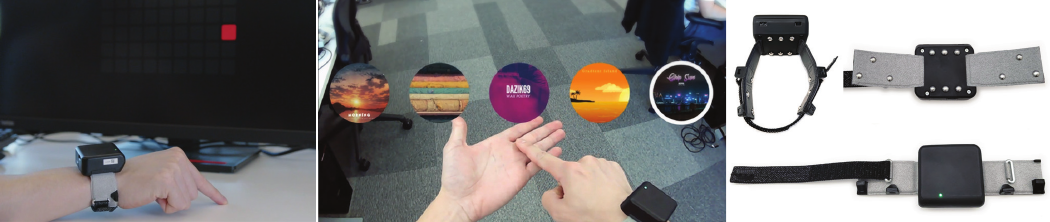}
  \caption{\sysname{} is a wristband that uses multimodal sensor fusion to unlock ubiquitous touch interactions. This enables touch interactions on environmental surfaces (left) and an always-available on-body interface (middle). The \sysname{} prototype (right) combines optical, electrical, and inertial sensing to derive estimates of touch state, touch gestures, and touch tracking.}
  \Description{}
  \label{fig:teaser}
\end{teaserfigure}


\maketitle

\section{Introduction} \label{sec:introduction}

With recent advances in both artificial intelligence (AI) and augmented reality (AR) technologies, constellations of mobile/wearable devices like smart glasses, wristbands, and smartphones are increasingly able to leverage context to digitally augment our physical interactions in the real world.
In our everyday lives, we interact with the physical world through touch, yet digital augments of physical objects (e.g., the ability to control an IoT device) are often locked away behind voice interfaces or smartphone menus.
We believe that ubiquitous touch interfaces are a critical part of an interaction model for future device constellations, but we lack a scalable, unintrusive, and socially acceptable way to sense our touch interactions with the real world.

Mid-air gestures, like pinch or thumb swipe microgestures, are commonly used as the core interaction model for egocentric AR systems, like HoloLens.
However, surface touch offers inherent tactile and proprioceptive feedback that allows for more confident eyes-free input.
Past work has found that compared to mid-air interactions, touch interactions leveraging physical surfaces improve performance on both manipulation tasks~\cite{cheng2022comfortable} and text entry tasks~\cite{dudley2019performance}.
The dominance of touchscreens in devices like smartphones and smartwatches underscore these advantages, but by breaking touch interactions out of the confines of small screens and into the environment (e.g., touch on your desk, on your palm, or on a handheld object) we can make these interactions more available and more contextually relevant.
There are also distinct advantages for \textbf{touch-on-world} (i.e., environmental surfaces), which can be highly adaptive to object identity and spatial context and can offer large interactive areas, and \textbf{touch-on-body} (e.g., the palm), which is always-available and less dependent on environmental/social context.

Past work has explored a variety of potential solutions to this longstanding problem, with significant focus on computer vision from a pair of glasses or a head-mounted display~\cite{richardson2020typing,xiao2018mrtouch}.
However, computer vision alone struggles at detecting precise fingertip contact with surfaces (i.e., the \textit{last-millimeter problem}), compelling users to exaggerate their motion, and touch interactions may not always be within the field of view of egocentric cameras.
Though there are many nuances of touch that enable new capabilities, we believe that the core building blocks for compelling use cases are: a) precise \textbf{touch state} detection - the ability to detect not just taps, but touch onset and offset, with a precision comparable to a capacitive touchscreen, 
and b) \textbf{touch tracking} - the ability to track fingertip position across a surface over time.
Ultimately, a compelling and complete touch interaction solution must be implemented with a single point of instrumentation in a socially acceptable form-factor that does not impede posture or dexterity.

In this work, we present \sysname{}, a wristband that uses multimodal sensing to enable touch interactions \textit{anywhere}.
We adopt a wristband form-factor as it is ideal for sensing hand activity~\cite{zhang2015tomo,ctrl2024generic,meier2021tapid}, is already proven as a socially acceptable form-factor, and does not require instrumenting the environment or concern about whether the hands are within the field of view of smart glasses.
\sysname{} aims to solve the core problems of touch state (for both in-world and on-body surfaces), 
and touch tracking for index-finger interactions.
Because these interactions involve complex muscle contraction, motion, and mechanical impact on physical surfaces, we adopt a multimodal sensor fusion approach that can capture the various phenomena involved.

Specifically, \sysname{} combines surface electromyography (sEMG), bioimpedance, inertial sensing, and optical sensing to recognize stateful touch, distinguish between conductive surfaces like the body and other environmental surfaces, and track both the fingertip and the hand over time. We train and validate our system on data collected from 100 participants and show that users can successfully complete various touch tasks, including a stateful touch task with a median success rate of 97\%, which points to the promise of this wristband sensing approach when scaled to a larger population.
Our multimodal approach makes \sysname{} a flexible research platform that enables a broad set of explorations into ubiquitous touch interactions and beyond.
As a research platform, we do not target a consumer-ready form-factor or optimize for power/compute, but prioritize diversity and robustness of sensing.

We envision that a wristband touch sensing solution like \sysname{} can complement smart glasses, augmented (AR) or virtual (VR) reality devices, or existing smartwatch interactions. 
To highlight the flexibility of this approach, in this work, all sensing is driven solely from \sysname{}, and other devices are used solely as examples of complementing \sysname{} with visual feedback.
In the following sections, we provide an overview of related work~(Sec~\ref{sec:related-work}), the sensing strategy and hardware implementation~(Sec~\ref{sec:system-design}), our inference pipeline for touch detection and tracking~(Sec~\ref{sec:touch-capabilities}), a collection of online evaluations~(Sec~\ref{sec:evaluation}), and a few interaction examples that offer a glimpse into how these capabilities provide value in a broader context~(Sec~\ref{sec:interactions}).

In this work our specific contributions include:
\begin{enumerate}
    \item A \textbf{multimodal wristband platform} optimized for touch interactions that combines sEMG, bioimpedance, inertial, and optical sensing
    \item Sensor fusion approaches that deliver core \textbf{touch primitives} including stateful touch and tracking on environmental surfaces. 
    \item A series of \textbf{interaction explorations} that leverage touch and touch context to highlight the potential of \sysname{} in future AI/AR constellations.
\end{enumerate}

\section{Related Work} \label{sec:related-work}
\sysname{} builds on a rich foundation of decades of touch research spanning interactions for multitouch displays, large-format displays, and augmented reality. Though \sysname{} targets similar touch capabilities, a major differentiation is \sysname{}'s ability to deliver these capabilities from just a single wristband, a prerequisite to operating in many mobile computing contexts. In this section, we contextualize our work with respect to past work on systems that enable touch by instrumenting the environment, wearable systems that enable touch interactions, and specific solutions to enable touch-on-body.

\subsection{Sensors in the Environment}
Direct instrumentation of surfaces is the most common approach to enable touch interactions and often sets the standard for accuracy.
Though capacitive, multitouch screens~\cite{lee1985tablet,hachet2011toucheo} offer unparalleled precision and are widely available, other efforts have sought to enable touch interactions on large, often rear-projected, surfaces using infrared illumination and cameras beneath the interaction surface~\cite{Rekimoto1997PerceptualS, han2005touch, holz2013fiberio}.
With the commoditization of depth sensing cameras, efforts like the work by Wilson et al. have pioneered the use of depth sensing to both detect and track touch on surfaces \cite{wilson2007depth,wilson2010lightspace,wilson2010depth}.
Xiao et al. improved touch performance by leveraging raw infrared images in addition to depth~\cite{xiao2016direct} and more recently, Shen et al. have used depth signal artifacts to improve touch precision at longer ranges, up to 3~m~\cite{shen2021farouttouch}. These direct imaging approaches excel at spatial tracking, but unless the cameras are placed behind the display, which is often impractical in everyday environments, cameras struggle to solve the "last-millimeter" contact problem. By placing the touch sensing in a wristband, \sysname{} avoids the need to add camera infrastructure to the environment.

To address this contact problem, the community has explored alternative imaging techniques that are directly sensitive to contact and force.
Laser ranging approaches (lidar) can be incorporated just above the touch surface, either on the surface itself~\cite{strickon1998laserwall} or incorporated in an object placed on a table~\cite{laput2019surfacesight}. When the finger crosses the plane of illumination and contacts the surface, light will be reflected back to the sensor.
ForceSight~\cite{pei2022forcesight} relies on the insight that most surfaces will deform slightly when touched and uses laser speckle imaging to estimate force from this deformation.
In some cases, thermal cameras simplify the problem of hand segmentation and tracking, but work like HeatWave~\cite{larson2011heatwave} and ThermoTablet~\cite{iwai2005thermal} leverage the heat transferred between the body and surface to detect and track touch interactions.

Other work has explored acoustic techniques that can turn a table or desk into a touch surface without the use of cameras.
When tapping on a rigid surface, acoustic waves propagate through the surface. Several efforts have mounted piezoelectric transducers at the corners of a surface and used arrival time and other signal characteristics to localize the touch location~\cite{ishii1999pingpongplus, paradiso2002knocks, pham2007acoustic}. 
Toffee~\cite{xiao2014toffee} made this approach more practical by placing the sensors directly on a mobile device, which rests on a rigid surface.
Others have shown how similar passive acoustic sensing techniques can be used to go beyond localizing taps and enable continuous or gestural input. Scratch~Input~\cite{harrison2008scratch} used a modified stethoscope to detect six different touch gestures on a table. TriboTouch~\cite{schultz2022tribotouch} used a micro-patterned surface to estimate fingertip velocity along a surface to improve touchscreen latency.

Acoustic techniques often impose the requirement that users tap with sufficient impact to be reliably detected, but capacitive touch interfaces more directly measure even light touches.
Projects like LivingWall~\cite{buechley2010livingwall} and Wall++~\cite{zhang2018wall} have attempted to capture the benefits of capacitive touch on large, wall-scale surfaces by embedding capacitive or resistive sensing in and behind paint.
Other efforts have brought capacitive sensing to everyday surfaces through rapid prototyping platforms that make it easy to instrument a surface~\cite{parilusyan2022sensurfaces,Pourjafarian2019MultiTouchKA,rekimoto2002smartskin}. Electrick achieved similar capabilities with relaxed surface requirements using electric impedance tomography and an easily applied conductive material. Touch\'{e}~\cite{sato2012touche} enables touch and interaction with physical objects using a swept-frequency capacitive sensing system attached directly to objects.

Instrumentation in the environment is an effective way to capture touch location, but these approaches do not scale to  use in new environments. In this work, we target ubiquitous touch interactions on any surface, which necessitates a wearable approach.

\subsection{Wearable Surface Touch}
With the rise in popularity of extended reality (XR) devices and smart glasses, a natural evolution for the imaging-based touch approaches is to move the sensors from the environment to a device on the head, where the egocentric perspective often offers an unobstructed view of the hands. Systems like OmniTouch~\cite{harrison2011omnitouch} and MRTouch~\cite{xiao2018mrtouch} use wearable depth sensing to estimate touch contact and location. Richardson et al.~\cite{richardson2020typing} enable text entry on environmental surfaces using VR-based hand tracking. TapLight demonstrates a wearable laser speckle imaging solution for both contact detection and tracking~\cite{streli2023taplight}. 
TriPad~\cite{dupre2024tripad} shares many of the same motivations as this work, and designs a variety of surface touch interactions using only existing hand tracking APIs provided by commercial headsets like the HoloLens 2 or Meta Quest Pro.
While these approaches effectively address portability concerns, many have the same \textit{last-millimeter} contact problem as environmental sensing since glasses-based hand tracking often has cm-scale fingertip positional error~\cite{soares2021accuracy}. 
Moreover \sysname{} seeks to enable touch sensing without requiring the user to wear glasses.

The most relevant body of work for \sysname{} is the use of wristbands or other hand wearables to enable touch interactions.
Passive acoustic sensing is the most deeply explored solution in this space.
WhichFingers~\cite{masson2017whichfingers} used piezo vibration sensors on each finger to recognize taps with finger discrimination, similar to commercial systems like the Tap Strap 2~\cite{tapstrap2}.
Ring-mounted IMUs have also demonstrated remarkable contact detection performance by leveraging the change in natural microvibrations when the finger is grounded against a surface~\cite{shi202readysteadytouch} or by classifying tap events~\cite{gu2019accurate}.
TelemetRing~\cite{takahashi2020telemetring} uses a similar sensing approach to detect when fingers are tapped, but the ring is made battery-free by relying on passive inductive telemetry to transmit data to an accompanying wristband.
MouseRing~\cite{shen2024mousering} uses one or two rings on the index finger and demonstrates remarkable fingertip tracking performance, comparable to a commercial touchpad.
VibAware~\cite{kim2023vibaware} uses an active acoustic transmitter on the wrist and a ring-mounted accelerometer to sense a variety of microgestures and touch interactions.

Systems like TapID~\cite{meier2021tapid} and TapType~\cite{streli2022taptype} demonstrate a more practical wristband form-factor using only inertial sensors to enable tap and text entry on any surface.
Acustico both detects and localizes taps using timing differences between surface waves and sound waves on wristband~\cite{gong2020acustico}.
Others have used sEMG measured with an armband for a more direct measure of contact force and finger discrimination~\cite{benko2009enhancing,becker2018touchsense}.
Optical flow sensors have been used to enable coarse tracking and hand-scale gestural input over surfaces~\cite{yeo2020wristlens}.

Despite all of these advances, there remains no wristband solution that solves for touch state, beyond simple taps. Though solutions exist for rings or glasses, we seek to enable this in a wristband form-factor for wider acceptability and without the field of view limitations of egocentric camera.
\sysname{} offers a more complete wristband research platform for touch interactions that includes stateful touch, not just taps. \sysname{} incorporates inertial sensing and extends sEMG sensing to a more practical wristband form-factor. We additionally add forward-facing optical sensors to track the hands and nearby surfaces and a novel bioimpedance solution to detect body-touch.

\subsection{On-body touch}
Finally, we consider approaches that specifically target on-body touch interactions. By projecting a touch interface on the palm, OmniTouch~\cite{harrison2011omnitouch} demonstrated compelling examples of always-available, contextually adaptive interfaces on the body, enabled by a wearable depth camera. 
A variety of sensing techniques have been considered to enable a more direct measure of finger-body interactions, including radar~\cite{hajika2024radarhand}, passive acoustic sensing~\cite{zhang2016tapskin,chen2019taprint,fathian2023face,Chen2020FaceOffDF}, active acoustic sensing~\cite{harrison2010skinput,liang2011sonarwatch,lin2011pub,mujibiya2013soundtouch}, and optical sensors around a wristband ~\cite{ogata2013senskin,ogata2015skinwatch, laput2014skinbuttons,xiao2018lumiwatch}.

The line of work on electrical touch detection is most directly relevant to our bioimpedance-based conductive touch module. For example, SkinTrack~\cite{zhang2016skintrack} uses a wristband sensor and ring-based transmitter to detect touch and coarse location by treating the body as an electrical waveguide.
ActiTouch~\cite{zhang2019actitouch} relaxes the form-factor requirements here by moving the ring to the sides of a VR head-mounted display.
ElectroRing~\cite{electroring} and Z-Ring~\cite{waghmare2023zring} use a similar technique, but demonstrates touch detection from a single point of instrumentation using multiple electrodes on a ring.
Other systems, like GestureWrist~\cite{rekimoto2001gesturewrist} or CapBand~\cite{truong2018capband} use a similar approach but focus on detecting wrist shape deformation, not body touch.
\sysname{} uses a two-row electrode configuration and high-sensitivity receiver to enable electrical touch detection anywhere on the body (on bare skin) or on external conductive surfaces from a single wristband, which represents a significant improvement in usability compared to past work.

\begin{figure*}[h]
  \centering
  \includegraphics[width=6in]{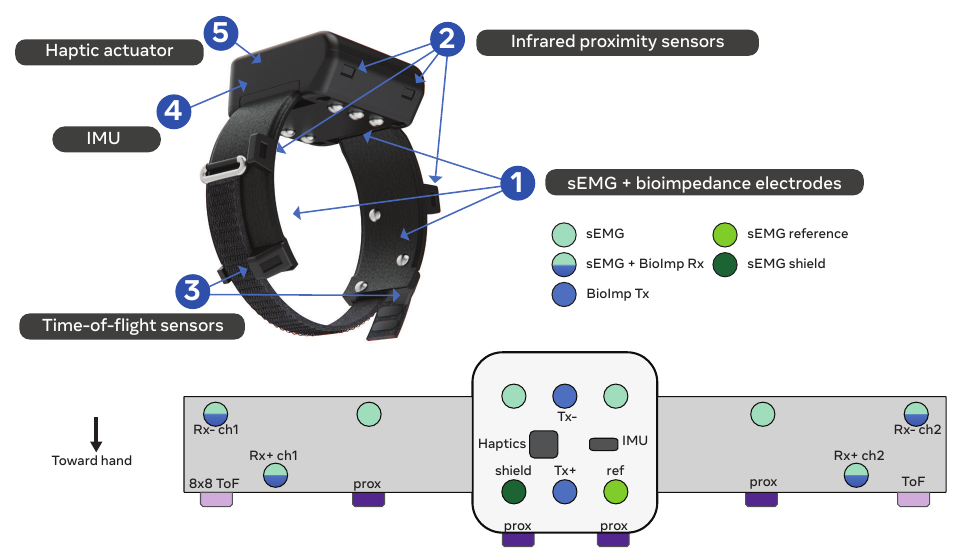}
  \caption{(top) A rendering of the \sysname{} device indicating each of the major subsystems with the puck and two active straps and (bottom) a diagram of the device as seen from below. (1) Electrodes distributed around the device measure sEMG and bioimpedance. Six shared electrodes around the strap measure both sEMG and bioimpedance signals in a time-multiplexed fashion. Beneath the puck are two additional sEMG electrodes along with the sEMG reference electrode, transmit electrodes for the bioimpedance signal, and an active shield electrode to reduce interference. (2)  Infrared proximity sensors around the strap measure deflection of the wrist and hand. (3) Time-of-flight sensors on the palmar side of the device track nearby potential surfaces beneath the hand. An IMU (4) and haptic actuator (5) are located in the puck with additional electronics.}
  \label{fig:sys-diagram}
  \Description{}
\end{figure*}

\section{System Design} \label{sec:system-design}

Even a simple touch interaction, e.g., tapping a virtual button, consists of a complex set of events. The hand moves toward the desired target and muscles in the forearm activate to flex the wrist and finger. The finger then makes contact with the surface, sending vibrations throughout the hand and potentially changing the electrical impedance of the hand. As the finger maintains contact and exerts pressure, the forearm muscles activate more strongly. Finally, as the finger releases contact, the wrist and finger extend and the hand moves back to a neutral position. A system that, e.g., only detects taps through vibration only captures a narrow slice of a touch interaction.

\subsection{Our Approach}

\sysname{} was designed from first principles to capture all of these facets of touch interactions. 
Specifically, we target two primary capabilities: (1) touch state, estimating whether the finger is currently in contact with a surface and (2) touch tracking, to estimate how the finger is moving along a surface.

\sysname{} captures both forearm motion and impact vibration~\cite{meier2021tapid} using low- and high-frequency signals, respectively, with an inertial measurement unit (IMU, accelerometer + gyroscope) augmented with a magnetometer.
When combined with kinematic constraints, we can use this to estimate hand position (relative to a stationary torso) and body pose using a technique similar to Shen et al~\cite{shen2016smartwatch}.
However, an IMU alone will tell us little about wrist deflection or touch-off events.
To compensate, \sysname{} also includes sEMG, which responds to wrist and finger motion as well as touch pressure.
sEMG contains significant information about hand motion, but incurs significant data and modeling complexity~\cite{ctrl2024generic}, so we complement this with optical proximity sensors to directly track wrist deflection using a technique similar to prior work~\cite{lee2011airtouch,salemi2021rotowrist,gong2016wristwhirl}.
For estimating touch state on bare skin on the body, we include a dedicated, bioimpedance sensor, which more directly measures stateful touch compared to the indirect inferencing required for touch on environmental surface with other modalities.
Finally, contextual information about surfaces near to the hand can greatly improve a model's ability to recognize motion as potential interactions with those surfaces.
Therefore, \sysname{} includes optical time-of-flight (ToF) sensors that detect and track nearby objects and surfaces.

Figure~\ref{fig:sys-diagram} shows the location of these modules in the \sysname{} prototype and Table~\ref{tab:sensors} summarizes each sensor. Our system consists of a central puck housing the battery and core electronics and two active straps, secured with a hook and loop. Refer to Appendix~\ref{sec:appendix-sys} for a full system diagram. In the following sections, we describe the included sensor modules in more detail.

\begin{table*}[h]
    \centering
    \begin{tabular}{rcccl}
         Sensor&  Channels&  Rate (Hz)& Max power (mW) & Measures\\
         \hline
         sEMG&  8&  2000& 144 & Forearm muscle activation\\
         Bioimpedance&  4&  2000& 972 & Electrical gradient along wrist\\
         Infrared Proximity&  4&  100& 29 & Wrist deflection\\
         ToF - single pixel&  1&  40& 75& Proximity to nearby surface\\
         ToF - 8x8&  64&  15& 244 & Contour of nearby surface\\
 IMU& 6& 890& < 3 & Acceleration, rotational velocity\\
 Magnetometer& 3& 110& < 3 & Magnetic field\\
    \end{tabular}
    \caption{\sysname{} sensor information. The max power listed represents the approximate power draw of each subsystem during a full-speed data collection.}
    \label{tab:sensors}
\end{table*}

\subsection{sEMG}
Surface electromyography (sEMG) is a sensing technique that measures electrical activity on the skin caused by a neuron's activation of a muscle. Often used in the medical field to assess muscle and nerve health, wrist- and forearm-derived sEMG has been shown to be a promising approach for gesture recognition~\cite{ctrl2024generic, saponas2009enabling}.  sEMG responds to changes in hand pose and force, necessary components of measuring touch interactions.

\sysname{} uses a custom 8-channel electrode configuration. The eight sensing electrodes are distributed around the wrist (three in each strap and two in the puck). As shown in Figure~\ref{fig:sys-diagram}, channels alternate between distal and proximal locations on the strap to maximize information content from different parts of the wrist. An active shield electrode in the puck helps reduce the impact of ambient electrical noise. sEMG channels are sampled simultaneously at 2~kHz using a dedicated analog-to-digital converter in the puck.

\subsection{Bioimpedance module}
\sysname{}'s novel bioimpedance sensor uniquely enables body touch interactions by detecting changes in impedance when the hand makes contact with surfaces of varying conductivity. Our solution is inspired by prior work~\cite{electroring,zhang2016skintrack,zhang2019actitouch,waghmare2023zring,waghmare2024zband} that couples a radio frequency (RF) signal to the body and measure the differential electric field across the skin. When the user touches a conductive surface, like the palm, the communication channel is altered, and this change can be detected by another pair of electrodes.

 \begin{figure*}[h]
  \centering
  \includegraphics[width=5in]{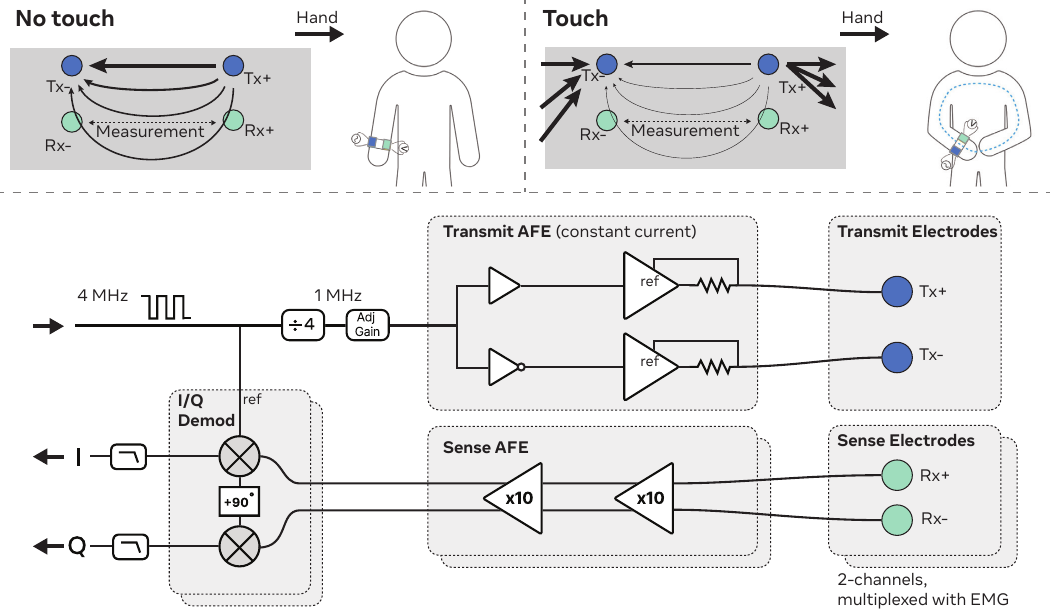}
  \caption{(top) Bioimpedance sensing principle for body touch: A constant-current differential transmitter (Tx) couples a low-current AC signal to the body. When touching the body, some of this signal is shunted off through this alternative current path. By placing differential sense electrodes (Rx) along the current path, we can measure the reduction in signal during a touch. (bottom) The bioimpedance module uses a differential transmitter operating at 1 MHz and two channels of differential sense electrodes that are processed with a synchronous detector.}
  \label{fig:bioimpedance}
  \Description{}
\end{figure*}

\begin{figure*}[h]
  \centering
  \includegraphics[width=5in]{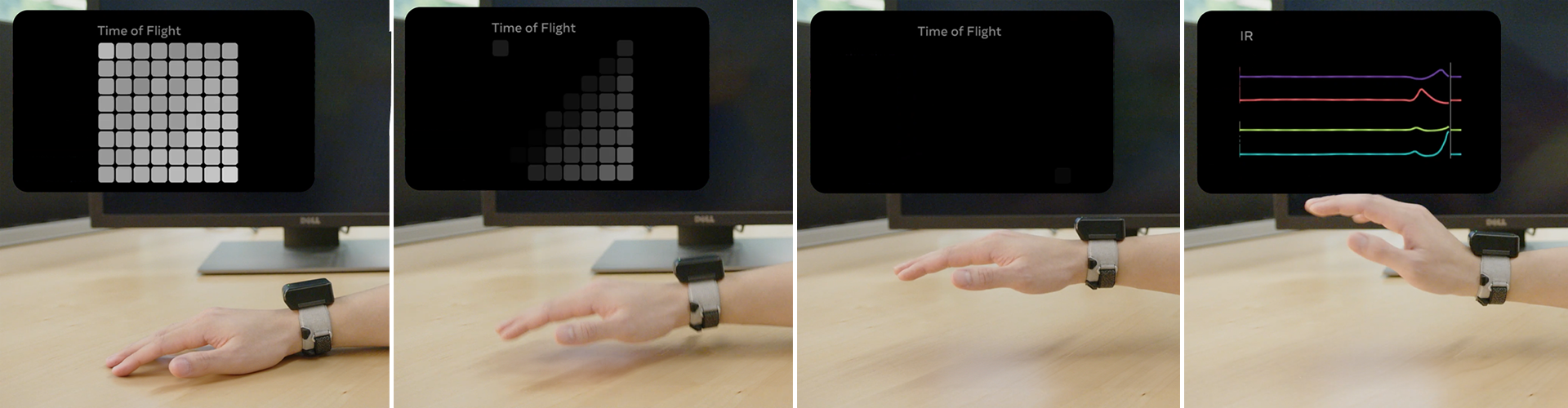}
  \caption{Example images captured by the 8x8 ToF sensor (left three images) and signals captured by infrared proximity sensors (far right). The ToF imager provides information about the hand pose relative to a nearby surface. Note that the sensor is offset from the middle of the wrist and positioned at an angle. Infrared proximity sensors around the top of the wristband provide information about wrist deflection.}
  \label{fig:proximity}
  \Description{}
\end{figure*}
Prior work that focused on touch detection relied on multiple points of instrumentation on the body~\cite{zhang2019actitouch,zhang2016skintrack} or used five rows of electrodes wrapped around the finger~\cite{electroring}, both of which impose significant encumbrance. To sense this from a single wristband while keeping only two rows of electrodes to minimize form-factor and encumbrance, \sysname{} uses two dedicated, differential transmit electrodes in the puck that galvanically couple a low-amplitude, high frequency, constant-current signal (1~mA at 1~MHz) to the body.

Unlike systems that require a transmitter on one hand and a receiver on the other, TouchFusion operates entirely from a single wrist. We measure the electrical field gradient within the wrist volume. When the closed loop is formed (e.g., touching the opposite palm), the impedance distribution across the arm changes, altering the local gradient detected by the receiving electrodes.

By measuring between a proximal-distal pair of electrodes, we observe a 1 MHz signal that changes slightly during a touch. To measure this subtle change 
we used a dedicated synchronous detector (AD8339) in the puck, which performs a quadrature demodulation, using the transmitted 1 MHz signal as a local oscillator, as shown in Figure~\ref{fig:bioimpedance}, bottom. This gives us in-phase and quadrature components at DC that we sample at a rate of 2 kHz.
Enabling these sensing electrodes to coexist with sEMG required careful simulation and design. Both paths share a common buffer and first stage amplifier. A multiplexer on each channel allows us to route the signal to either the sEMG signal path or the bioimpedance path. We performed additional amplification and filtering along the bioimpedance path.

\subsection{Optical Proximity \& Time-of-Flight}
In addition to the \textit{internal} sensing afforded by EMG and bioimpedance, \sysname{} leverages \textit{external} optical sensing to directly capture wrist deflection and to detect nearby surfaces.
Optical sensing offers a uniquely complementary signal with less variance that is less sensitive to motion artifacts. \sysname{} includes three distinct types of optical sensors. Four infrared (IR) proximity sensors (VCNL4020) -- two in the puck, one in each strap -- directly respond to the distance between the band and the hand. As the wrist deflects, the skin reflects different amounts of infrared light to each sensor. The light is modulated at 390 kHz to reject ambient illumination. Due to the placement closer to the dorsal side of the hand, this configuration is most sensitive to wrist extension and ad-/abduction.

On the palmar side of the band, \sysname{} includes two infrared time-of-flight (ToF) sensors. One is a single-pixel optical rangefinder (VL6180X) and the other is an 8x8 multi-zone ToF sensor (VL53L5CX). The single-pixel sensor is most useful for easily sensing proximity to a surface, allowing \sysname{} to dynamically adapt operating modes, e.g., to enable a particular inference model when a surface is detected beneath the hand. Because this sensor has a narrow field-of-view and has no way of distinguishing between hover over a table or any other occlusion, the 8x8 multi-zone ToF sensor allows us to estimate a coarse "map" of the nearby surface, classify whether it is planar, and detect edges, as shown in Figure~\ref{fig:proximity}.
Each of these optical sensors is self-contained and digitally communicates with the wristband puck. Care is taken to shield digital communication traces from sensitive sEMG traces in the strap.

\subsection{Integration}
In the puck, \sysname{} includes a sensor aggregation board that handles sEMG and bioimpedance analog processing and digital communication. It additionally includes a haptic linear resonant actuator (LRA) for vibration-based haptic feedback. The sensor aggregation board mates to a general-purpose compute board that includes a PSoC 6 microcontroller, (CY8C6347BZI) a BLE antenna, an IMU (LSM6DSRXTR), a magnetometer (LIS2MDLTR), a USB-C port, and battery management circuitry. The device is powered by a 950 mWh battery and lasts approximately 4 hours with a profile consistent with daily use or 45 minutes with most subsystems powered on at full speed (this configuration was used for data collections and demos).

The straps of the device consist of a flexible printed circuit with stainless steel electrodes. The strap is encapsulated in a textile laminate for durability and comfort. To accommodate a diverse population, we designed two sizes of the device intended to cover the 3rd to 91st percentile of wrist sizes. However, we note that users with wrist sizes near each of these boundaries will face degraded sensing due to shifted sensor locations. 
We use a hook-and-loop strap that wraps around the active straps for donning.

\begin{figure*}[h!]
  \centering
  \includegraphics[width=5in]{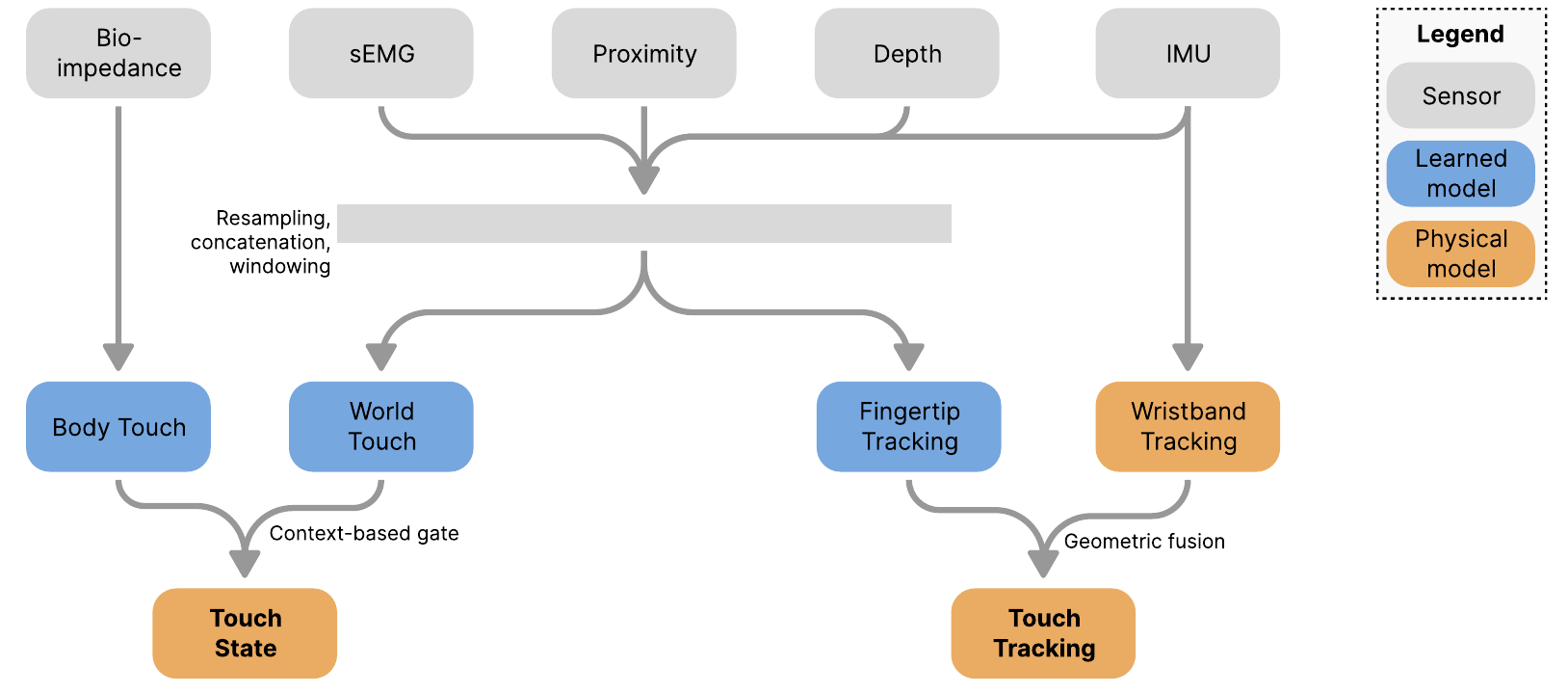}
  \caption{\sysname{} estimates touch state and tracks motion during touch events. We decompose this problem into smaller independent modules in a late-fusion approach. Each of the sensors in \sysname{} was chosen to contribute to one of these modules. This multimodal approach contributes to the breadth of \sysname{}'s capabilities and improves resilience.}
  \label{fig:model-arch}
  \Description{}
\end{figure*}

\subsection{Online pipeline}
For data collection and interactive applications, \sysname{} communicates over BLE with a host PC running our online pipeline. We adopt a modular, graph-based pipeline architecture that allows us to asynchronously run multiple preprocessing steps, algorithms, or model inferences simultaneously. For data collection, we disable all inference and simply record all raw data in the HDF5 format. For interactive applications, we enable all algorithms and models in a single pipeline. Though no attempt has been made at model optimization, we note that all models described in this paper can run simultaneously on a desktop PC with an Nvidia RTX 3080 GPU and AMD Ryzen Threadripper PRO 3975WX.

\section{Modeling Approach} \label{sec:touch-capabilities}




To enable the two primary interaction targets – Touch State and Touch Tracking – we adopt a fusion approach, first combining subsets of our data streams to infer semantically meaningful intermediate quantities (e.g. forearm motion), and then performing a fusion of those inferences to yield a final interaction model. Briefly, those intermediate inferences include:

\begin{itemize}
    \item Body Touch – Uses bioimpedance information to infer body self-contact, such as when the finger is in contact with the opposing palm.
    \item World Touch – Classifies touch onset and offset events on surfaces, from which an estimation of touch state can be derived for a given point in time.
    \item Fingertip Tracking – Estimates the position of the fingertip relative to the wrist, as projected onto an XY plane beneath the user’s hand.
    \item Wristband Tracking – Estimates the user’s wristband position relative to the body in 3D space.
\end{itemize}

Figure~\ref{fig:model-arch} shows an overview of the full pipeline, highlighting how each sensing modality is used in downstream models. 
Because the different sensor types are sampled at different rates, we first simplify the training of any learned models by resampling and time-aligning (via interpolation) all data streams to our highest sampling rate (2~kHz). These data streams are then concatenated and windowed at a fixed-length time step (variable per model). As a result, during online inference, we process incoming data using a sliding window, and perform inference on a frame-by-frame basis at a target inference rate. 
Section~\ref{sec:datasets} outlines our methods for collecting and labeling data, and the subsequent sections describe our implementations for each model in the pipeline.

\subsection{Datasets}
\label{sec:datasets}
To develop these models, we collected data from 100 participants recruited from the community surrounding our institution. Each dataset, summarized in Table~\ref{tab:datasets} consists of an approximately 1 hour data collection protocol during which participants wore the \sysname{} device and performed a variety of prompted touch interactions. Participant actions were prompted using a monitor on the desk. Participants were compensated with a \$75 gift card. To support the breadth of touch primitives, we collect two variants of this protocol, described in the following sections. All interactions were logged and synchronized using a nearby desktop PC. 


\subsubsection{World Touch Dataset}
This dataset supported most of the model development and consists of N=70 sessions from unique participants. 
During a data collection session, participants sat at a desk and touched on a Sensel Morph touchpad~\cite{sensel}, which logged contact, force, and position information to provide ground truth annotations.
The touchpad, the \sysname{} device, and the user's index finger were instrumented with small retroreflective fiducial markers, and OptiTrack cameras around the desk tracked participants' hand motion.
Prompted actions included taps, swipes, and touch \& hold for specified durations.
To train the fingertip tracking model, we prompted target acquisition and drawing of specified shapes to elicit a variety of continuous pathing activity.
Crucially, to ensure robustness in real-world usage, we explicitly trained on 'null class' data. This ensures the model discriminates intentional touch interactions from common daily motions. To collect null class data, we prompted users to "point" at a target in the air using their wrist and prompted a variety of freeform motions, including claps, pinches, waves, etc.
 

\subsubsection{Body Touch Dataset}
We collected a second dataset (N=30 participants) to develop our body touch module. Each session contained 40 minutes of index finger touch on the opposing palm, including taps, press and hold, and swiping/drawing on the palm. We include simple null classes like rotating the wrist, swinging the arm, and "fake" taps where one pretends to touch the opposing palm without actually making contact.

Collecting reliable annotations for body touch posed a challenge, as direct instrumentation of the touch point on the body would change the behavior of the bioimpedance module. Instead, we used a second \sysname{} device and invited participants to wear one on each wrist. The device on the dominant hand functioned as a standard \sysname{}, with all subsystems enabled, while the one on the non-dominant hand had the RF transmission disabled. Much like in SkinTrack~\cite{zhang2016skintrack}, we used the bioimpedance receive signal on the non-dominant device to annotate touch events. The data from this secondary device was only used for ground truth annotation and was not used as training data.

\begin{table}
    \centering
    \begin{tabular}{r|C{3 cm}|C{3 cm}}
         &  World Touch& Body Touch\\
         \hline
         \# subjects&  70& 30\\
         Age&  35.2 ($\sigma=9.5$)& 38.5 ($\sigma=12.6$)\\
         Gender&  39 M, 27 F, 4 other& 12 M, 18 F\\
         Handedness&  64 R, 6 L& 28 R, 2 L\\
 Device size& 24 L, 46 S&19 L, 11 S\\
 \hline
 Used for& world touch state, touch gestures, fingertip tracking&body touch state\\
    \end{tabular}
    \caption{Summary of the two datasets we collected to train our touch models.}
    \label{tab:datasets}
\end{table}

\subsection{Touch State}
We model touch state as distinct onset and offset events, as these are usually more discernible compared to touch and no-touch states and it is easy to miss a brief touch state. 
We define our null class as anything other than onset/offset events.
We develop two variations of this model---one optimized for touch on environmental surfaces, \textit{world touch}, and one that is driven by the bioimpedance sensor to detect \textit{body touch}.

\subsubsection{World touch model}
For each windowed event, we consume sEMG, the four channels of IR proximity sensing on the top and sides of the band, and IMU signals to form the feature set. We use a sliding window of 1~s length with a stride of 8.5~ms.
A CNN-LSTM model with max-pooling is used for classifying null, onset, and offset events. For full details on model architecture, refer to Appendix~\ref{sec:appendix-models}. The convolutional layers extract time- and frequency- domain features from the input window, and the LSTM layer learns feature relationships over longer time scales.
This model was trained using the world touch dataset on blocks that include longer pathing motions like drawing shapes or acquiring a target.

\subsubsection{Body touch model}
For body touch, we aim to support highly subtle touches, suitable for use when rapidly swiping over the palm. Because of reduced finger motion and vibration involved here, we use only the 2-channel complex bioimpedance signal with a sliding window of 250 ms and a stride of 6 ms. 
We again use a CNN-LSTM architecture, consisting of alternating 1D-convolution and max-pooling layers, followed by a single LSTM layer, as detailed in Appendix~\ref{sec:appendix-models}.
This model was trained using the body touch dataset using study blocks that included tap and swipe gestures on palm.

\subsubsection{Evaluation}

\begin{figure}[h]
  \centering
  \includegraphics[width=\linewidth]{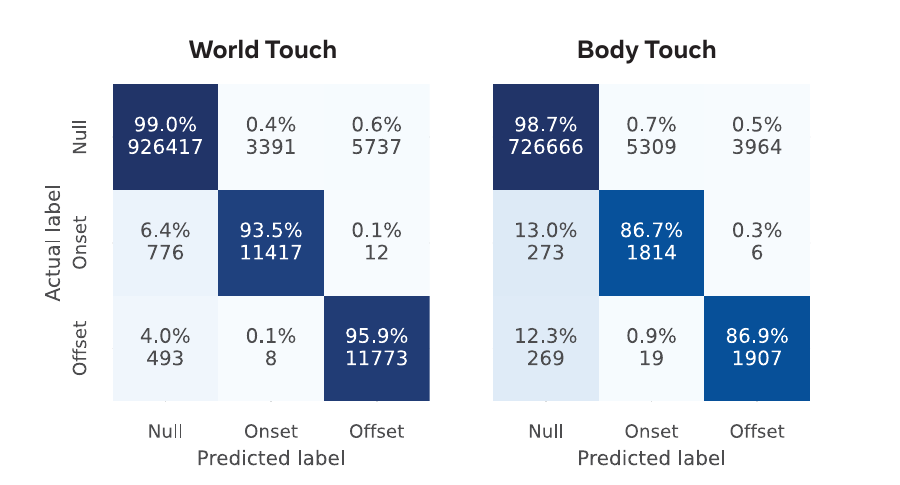}
  \caption{Confusion matrix for stateful world touch (left) and body touch (right) models. Results are reported on a per-window basis.}
  \label{stateful-touch-conf-matrix}
  \Description{}
\end{figure}

Each model was evaluated using 10-fold cross-validation across users, measuring accuracy in the predicted class on isolated windows of data. Window-level confusion matrices for each model are shown in Figure~\ref{stateful-touch-conf-matrix}. The world touch model achieves AUC-PR of .960 and AUC-ROC of .994, while body touch achieves AUC-PR of .794 and AUC-ROC of .969. During inference, we apply postprocessing that includes a configurable per-class threshold and debouncing (i.e. temporary disabling of repeat predictions) with values of 160~ms and 56~ms, for the on-world and on-body models, respectively. This allows us to predict discrete events as they occur with higher confidence. See Section~\ref{sec:evaluation} for an evaluation in a real-world setting.

Although these metrics are not directly comparable since the models were trained using different datasets, the results seem to suggest that the multimodal world touch model performs substantially better than the bioimpedance-driven body touch model. We attribute this difference to: 1) body touch often involves more subtle finger motion, less impact, and more arm motion compared to touching on a rigid surface, making this a more challenging task. 2) The bioimpedance signal is dependent on a stable skin-electrode interface, and poor contact (e.g., due to wrist anatomy variation) leads to unusable output. 
In practice, we observe that body touch performs reliably once the device is appropriately tightened and the skin-electrode interface is stable. This underscores the importance of future work to improve band/electrode compliance and to fuse bioimpedance with modalities like EMG to improve performance.

\begin{figure*}[!h]
  \centering
  \includegraphics[width=6in]{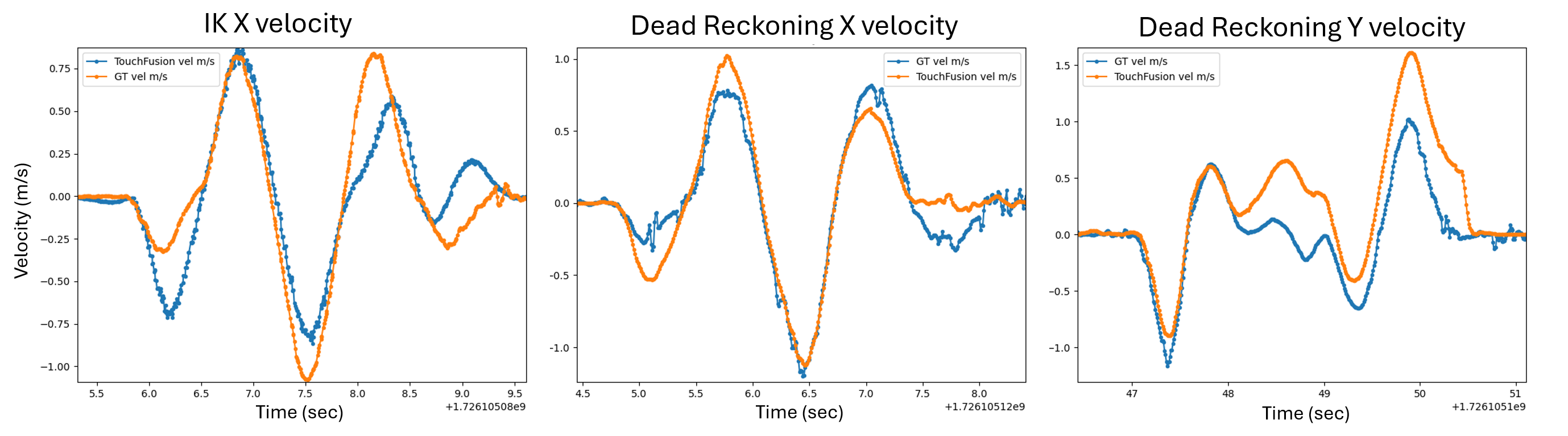}
  \caption{Representative snippets of velocity traces for each of our wrist tracking models compared to ground truth from OptiTrack. Traces are extracted from P8, whose tracking performance was close to the population mean (IK X: \textit{Mean Error} 24.4cm/s, \textit{RMSE} 35.3cm/s, \textit{R-score} 0.81, Dead Reckoning X: \textit{Mean Error} 15.2cm/s, \textit{RMSE} 20.8cm/s, \textit{R-score} 0.90, Dead Reckoning  Y: \textit{Mean Error} 20.9 cm/s, \textit{RMSE} 29.9 cm/s, \textit{R-square} 0.89)}
  \label{fig:tracking-p8}
  \Description{}
\end{figure*}

\subsection{Touch Tracking}
Tracking touch location requires both coarse wristband tracking - i.e., where is the wristband relative to the body -- and fine-grained fingertip tracking -- i.e., where is the fingertip relative to the wristband.

\subsubsection{Wristband Tracking}

We use the IMU for coarse wristband (forearm) tracking. To localize the wristband with respect to the body, during active use we assume the user position in the world to be stationary and the body upright. We use a standard 3-DoF filter to estimate orientation from the IMU and magnetometer, and then apply an orientation-constrained inverse kinematics (IK) model, similar to~\cite{shen2016smartwatch}, to estimate the dominant arm's pose. Moving the arm left/right generally induces changes in the arm orientation, but this approach is unable to capture motion in the forward/backward direction. For applications that require only 1D tracking, we can opt to use the IK estimate directly for drift-free tracking, but for 2D applications, we designed a dead reckoning algorithm to track the wristband along a surface while minimizing drift.

We first use the arm orientation to derive a gravity-subtracted estimate of acceleration, which we double-integrate to infer position. Though trackpad-like interactions with closed-loop feedback are more sensitive to velocity error than absolute position error, the drift from a naive dead-reckoning approach is still insufficient for reliable control due to even small bias in the acceleration measurement.
To correct for this, we apply two techniques to mitigate drift: 1) \textbf{Stationarity detection:}  When the variance and magnitude of the acceleration and rotational velocity over an interval are sufficiently small, we determine that the wristband is likely stationary and exponentially decay the velocity estimate to zero. 2) \textbf{Bias compensation:} During periods of continuous motion, constant velocity motion across a single axis is unlikely to occur due to kinematic constraints. Over short intervals, we assume that the acceleration bias is constant, so we approximate it as linearly increasing error in the velocity estimate. Using linear regression, we compute a running estimate of the acceleration bias on each axis and subtract it from our velocity estimate with a small gain.

\subsubsection{Fingertip Tracking}
The fingertip tracking model estimates the position of the user's index fingertip relative to the wrist, which includes the two degrees of freedom each of the wrist and metacarpophalangeal joints as well as the two additional degrees of freedom from interphalangeal flexion/extension.
\sysname{} lacks a direct measure of these degrees of freedom, so we trained a multimodal regression model that estimates 2D fingertip velocity along a surface. This design helps to improve the generalizability of our population model, as the relative changes in our signals from motion generalize better than absolute values corresponding to pose, which vary greatly across hands.
Specifically, this model operates on fused sEMG, IR proximity sensors, and IMU signals over a 200~ms window with a stride of 12.5~ms. IMU and IR pass through depthwise-separable convolutional layers before concatenation with covariance features from the sEMG data. This joint feature set is fed through two linear layers to obtain the final regression.

This model is trained with Sensel touchpad ground truth annotation from the surface touch dataset.
We perform a rotation of the Sensel motion vectors based on hand and arm pose from  OptiTrack, to correct for the fact that participants do not necessarily approach the Sensel straight-on. 
We introduce a number of post-processing optimizations to make these raw velocity estimates usable, including smoothing with a OneEuro filter and an adaptive baseline approach that suppresses drift when not in motion. 
Specifically, if we are in a sufficiently long "stable region" of limited motion, i.e. both the norm and the variance of the X/Y signal remain below target thresholds, we set the new baseline to be the mean of the stable region and subtract this baseline from subsequent values.
We suppress remaining jitter/drift by setting any sample to (0, 0) if the signal norm remains sufficiently close to the baseline. 
For interactive applications, we obtain a final position estimate of the fingertip by fusing the wristband velocity with fingertip velocity and integrating the result.

\subsubsection{Evaluation}
We evaluate the performance of our touch tracking models by comparing estimated wrist and fingertip velocities to ground truth velocities derived from OptiTrack or Sensel, respectively.
Because the wrist tracking model is not data-driven, we collect a supplemental evaluation dataset from additional 8 right-handed participants (5M, 3F) with ages ranging from 25-30.
Participants wore a \sysname{} device instrumented with fiducial markers and sat at a table with additional markers to assist in the alignment of coordinate frames between the \sysname{} device and OptiTrack.
We first asked participants to use whole-arm motions to trace out a star pattern between five marked points. This was repeated 20 times with a 1-2 second pause between each trace.
OptiTrack captured position at a frequency of 120~Hz, and we recorded the estimated velocity from both the IK and dead reckoning wrist tracking algorithms at \char`~ 100Hz. 
Due to technical issues during the study (e.g., device battery low), data collection was repeated for three of the participants.

\begin{figure*}[h]
  \centering
  \includegraphics[width=6in]{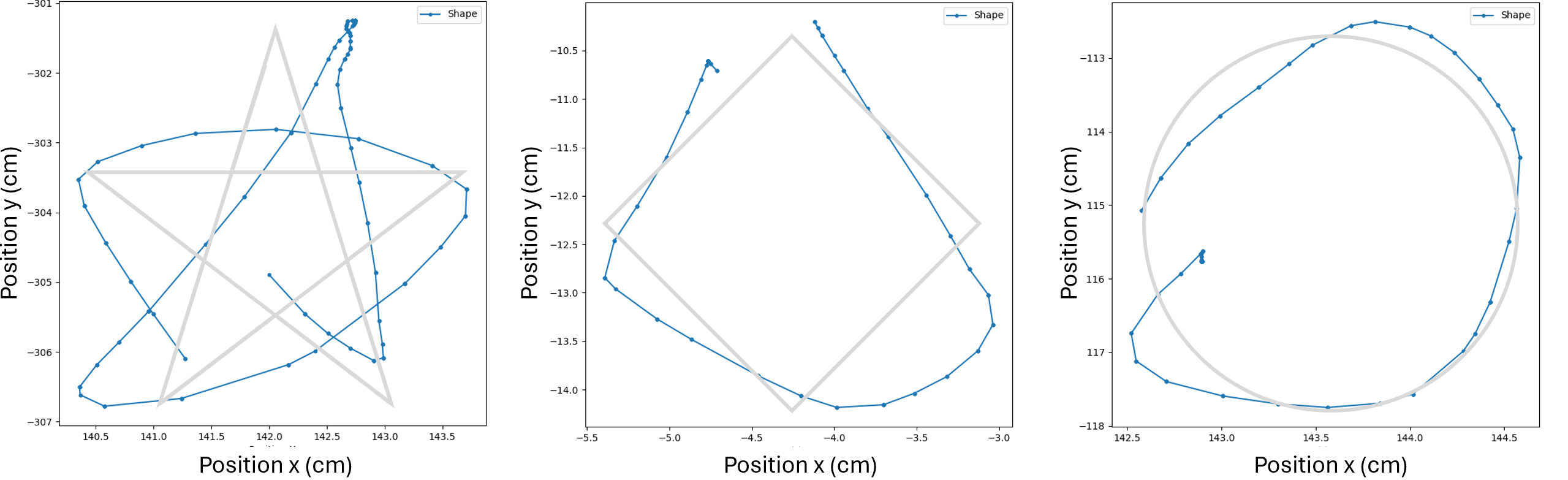}
  \caption{Representative shape traces derived from the post-processed fingertip model.}
  \label{fig:finger_traces_okay}
  \Description{}
\end{figure*}

To evaluate our wrist tracking algorithms, we derive the ground truth wristband velocity by taking differences in OptiTrack poses across multiple samples to reduce noise.
We compare this to the smoothed 1D velocity (x) derived from the IK model and the 2D velocity (x, y) derived from the dead reckoning wrist tracking algorithm.
The results are shown in Table~\ref{tab:tracking_results}.
Figure~\ref{fig:tracking-p8} shows a representative example of x and y velocity traces compared to ground truth for both the IK and dead reckoning approaches from P8, whose metrics are similar to the population mean.
In general, we observed more velocity drift from the y (depth) axis compared to the x axis.
The error in both axes are significant, though unsurprising given the integration of acceleration.
However, we note that that in a closed-loop interaction where the user receives visual feedback, this still allows some degree of user control.

\begin{table}
    \centering
    \begin{tabular}{r|c|cc|c}
         Solution &  IK (1D)&  \multicolumn{2}{c}{Dead Reckoning} &Fingertip Tracking\\
         Axis&  x&  x& y &x+y\\
         \hline
         Avg Speed (cm/s)&  24.4&  24.4& 22.4 & (unavailable)\\
         Mean Error (cm/s)&  12.7&  14.9& 17.7 &3.5\\
         RMSE (cm/s)&  19.4&  23.2& 26.3 &5.8\\
         R&  0.86&  0.86& 0.84 & 0.95\\
    \end{tabular}
    \caption{Tracking results for our wrist and fingertip tracking approaches. Wrist tracking results were computed on a separate N=8 dataset. Fingertip tracking results are cross-validated results from the surface touch dataset.}
    \label{tab:tracking_results}
\end{table}

For fingertip tracking, we evaluate the model using 10-fold cross validation across users in our surface touch dataset, using study blocks that contained pathing trials.
The learned model achieves an R-squared score of .904, before any additional post-processing. 
Figure~\ref{fig:finger_traces_okay} shows examples of representative traces from several users. Please refer to Appendix~\ref{sec:appendix-study} for additional best/worst-case examples. Note that these traces were performed open-loop by tracing paths on a printed sheet of paper on the table. No additional visual feedback was provided.

\subsection{Additional Functionality}
An expressive touch interaction language includes not only continuous touch input, but also discrete gestures, like taps and directional swipes.
While we could derive classifiers for these gestures using the outputs of our touch state and touch tracking models, we additionally trained an explicit lightweight recognizer for these common gestures to run in parallel with our state/tracking models – or in place of them when only simple, discrete UI navigation is needed.
We use a similar model architecture as the world touch model, but using only IR and IMU data streams for simplicity.
This discrete gesture functionality has been shown in previous wristband devices~\cite{kim2023vibaware,tapstrap2,meier2021tapid}, so we do not claim this as a contribution, but we include this functionality as a building block for our interaction explorations in Section~\ref{sec:interactions}.

\section{Online Evaluation} \label{sec:evaluation}
To supplement our promising model-level metrics in the previous section, we additionally assessed a subset of our touch primitives in standardized online tasks, where users receive immediate visual feedback.
For learned primitives, these model outputs are post-processed, unlike the window-level metrics previously reported.

Fifteen right-handed participants were recruited from the community to participate in the study. Participant ages ranged from 25 to 63 (mean age = 33.3 years, SD = 11.4 years). By gender, 6 participants each identified as male and female, while 3 participants identified as non-binary. Participants were provided a gratuity exchangeable for corporate merchandise as compensation for an hour of their time.

\begin{figure*}[h]
  \centering
  \includegraphics[width=5in]{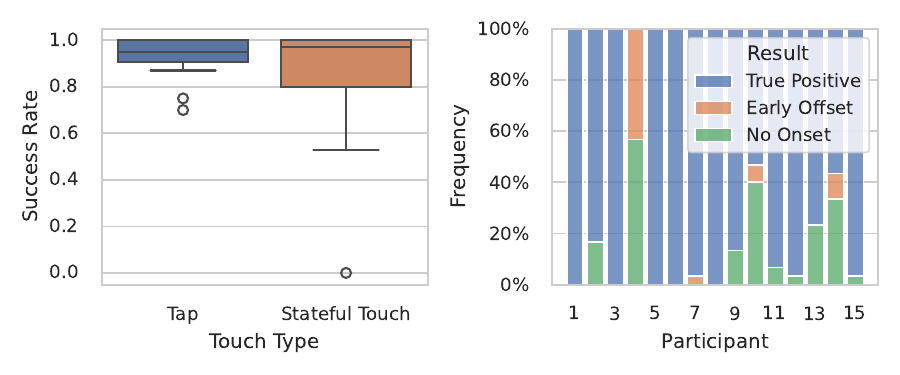}
  \caption{(left) Success rates for Tap versus Touch and Hold for the Dialog Dismissal task. (right) Breakdown of failure types for Touch State}
  \label{fig:user-study-results}
  \Description{}
\end{figure*}

Participants were welcomed, seated in a swivel chair at an office desk, and asked to review and sign a form to confirm their consent. Participants could rest their elbow on the arm of the chair if they wished. The study administrator then helped the participant don the wristband, and the study began. 
The study proceeded in multiple phases presented in random order; each phase featured a single task that evaluated a specific type of interaction. In this work, we considered three types of touch primitives: a basic tap gesture, stateful world touch, and 1D touch tracking. In each phase, the participant was provided video instruction on the action that they would be required to perform and asked to practice along with the video. A second video introduced the task they would use this action to perform, one of: dialog dismissal or a Fitts' task. The study then proceeded with the test trials for that task, followed by a questionnaire that collected subjecting ratings from 0-10 on a variety of standard experiential characteristics for that input technique.
The following sections describe the two task types followed by the results by touch primitive.

\subsection{Evaluation Tasks}
\subsubsection{Dialog Dismissal Task - Tap and Touch State}
We evaluate the tap gesture model and the on-world touch state model with a task that simulated dismissing a dialog box.
The task for stateless tap consisted of a popup that required the user to perform a surface tap within 2 seconds of presentation. For stateful gestures, the dialog box included a progress bar indicating the amount of time the participant needed to hold a surface tap for the gesture to be considered successful. The required hold time varied randomly between 0.5 and 3 seconds. This evaluation consisted of 6 blocks of 6 dialog box trials, for a total of 36 gestures issued, followed by a subjective questionnaire. The time between trial presentations was randomly chosen between 1 and 3 seconds to reduce participants' ability to anticipate the prompt.

\subsubsection{1D Fitts' Task - Continuous Touch Tracking}
To assess the touch tracking capabilities of \sysname{}, we developed a 1D Fitts' Law-based reciprocal selection task~\cite{fitts} based on the fused wristband + fingertip tracking. In the task, users moved a cursor between two circular targets, in succession, for 9 selections, finalized using a dwell time of 250 ms. 
The Fitts' sub-study was a 2-width (small = 7mm, medium=9mm) x 2-amplitude (near = 45mm, far = 79mm) x 2-direction (up-down, left-right) x 4 block within-subjects design, covering indices of difficulty (ID = $\log_2{2A/W}$) from 3.3 to 4.5.

Our decision to use a 1D evaluation task was based on two factors. First, it allowed us to separately analyze the Up-Down and Left-Right axes of motion, which we hypothesized may have meaningful differences in performance. Secondly, we observed a larger learning curve when participants first encounter precise 2D targeting with \sysname{}. Because our current model does not personalize for each user, participants instead adapt over time the direction, force, and speed with which they control a cursor until they become proficient. At the same time, we found that calibration of postprocessing parameters was effective in remapping their actual motion to an intended XY orientation, given any differences in how the device was worn. In contrast, we found 1D motion to be intuitive and afford confident control right away, and we therefore limited the task to 1D to reduce the need for training and calibration.

\subsection{Results}
\subsubsection{Tap and Touch State Results}
Of the six blocks of this task, we considered the first block as training, and report metrics from blocks 2-6, or 30 trials per participant per gesture. During analysis, we observed some data loss in the logging for tap trials that affected all six blocks for P7 and three blocks from P8. This does not affect touch state analysis.

Overall we found a median success rate of 0.95 (mean=0.93, SD=0.1) for tap gesture trials and 0.97 (mean=0.83, SD=0.28) for touch state trials, as shown in Figure~\ref{fig:user-study-results}, left. To better understand the types of errors experienced for touch state trials, we analyzed the occurrences of touch onsets and touch offsets within the trials and found that failures largely stemmed from the model failing to detect the touch onset (15\% of total trials and 76\% of unsuccessful trials), although one participant (P4) was particularly impacted by early release errors, accounting for nearly half their trials (43\%) (Figure~\ref{fig:user-study-results}, left).

The surveys of participants' subjective experience showed that participants found the Tap and Touch State gestures to be highly satisfying, supported high feelings of success, and were deemed highly learnable. Participants also rated the interactions extremely low on discomfort, fatigue, frustration, mental demand, and physical demand (see Appendix~\ref{sec:appendix-study} for full results).

\subsubsection{Continuous Touch Tracking Results}
We calculated the Fitts' throughput metric directly using guidance from Mackenzie's recommended calculation for reciprocal tasks~\cite{fitts}, which computes the effective movement distance and effective target width directly for each trial. This calculation was performed on each trial, consisting of 8 target selections, then averaged across all trials.
Overall, users had consistently better performance in the Left-Right direction than the Up-Down direction, with higher throughput rates (1.23 bits/s) and lower reentries per target (0.21) than Up-Down, which had throughput of 1.09 bits/s, and reentries of (0.53) per target. Taken from literature on extensions to Fitts’ metrics~\cite{mackenzie2001accuracy}, reentry is the number of expected times a user’s cursor enters the target bounds after the first entry, and is an indication of the stability of pointing. 
While the throughput numbers are relatively low, they are on par with nascent input systems such as gaze, and likely reflect the relatively high ID of the tasks in all conditions. Subjective responses to a 10-point NASA-TLX inspired survey found that users agreed that Left-Right outperformed Up-Down, reporting that it provided higher satisfaction (median 8 vs 5), success (median=8 vs 7), and learnability (8 vs 7), and was rated considerably lower on fatigue (median=0 vs 3) and frustration (median=1 vs 5) than Up-Down. See Appendix~\ref{sec:appendix-study} for more details on subjective measures.

\section{Interaction Explorations} \label{sec:interactions}
To showcase the value of these touch capabilities, we present a series of interaction explorations built on \sysname{}. These demonstrate how the touch primitives we've developed can be integrated in a real application and provide building blocks toward a touch-based interaction language. All interactions presented were built with a \sysname{} device wirelessly connected to a desktop PC. Visual interfaces were implemented on the PC in a Unity application and streamed to an off-the-shelf lightweight smart glass display (LetinAR T-Glasses).

\subsection{Summoning a Ubiquitous Control Surface}
\begin{figure}[h]
  \centering
  \includegraphics[width=\linewidth]{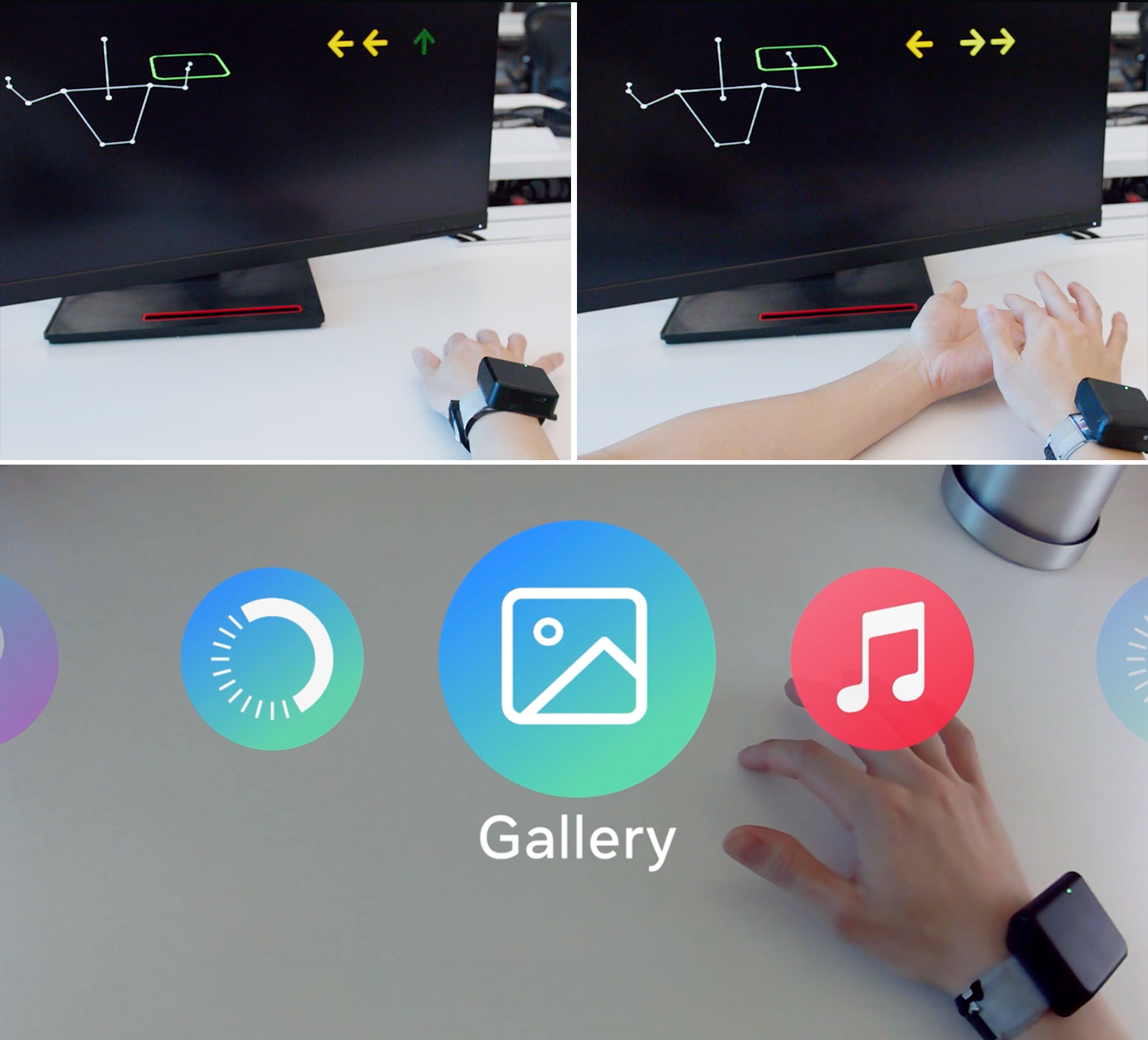}
  \caption{By double tapping a nearby surface like desk (left) or the opposing palm (right), a user can summon a control surface that enables taps and direction swipes. The control surface is anchored to the tap location and deactivates when the hand leaves the control surface.}
  \label{fig:summoning}
  \Description{}
\end{figure}

An ideal AR/AI system remains transparent to the user as they go about their world, until the user wishes to interact with it. We demonstrate a simple interaction model built with \sysname{} that allows the user to invoke an otherwise ambient AR interface by simply double-tapping on their palm, leg, or a nearby surface.
\sysname{} uses the downward facing optical sensors to detect that the hand is hovering over a plane. If not, we disable activation to prevent false positives. As shown in Figure~\ref{fig:summoning}, we track the wearer's upper body pose upon summoning to detect when they are within bounds of the control surface - defined as within approximately 6 inches of their initial touch point. If the hand leaves the control surface for more than a few seconds, it deactivates. This illustrates two gates that help prevent false positives: 1) the user's hand must be above a horizontal planar surface (this could be extended to work with other types of surfaces), and 2) the user must double tap to activate the system.

We can use this control surface to interact with interfaces through AR glasses, a smartwatch display, or an external screen in the environment (e.g., a television). In this example, we show how basic gestures (tap and swipe) can be used to control a simple linear app interface~(Figure~\ref{fig:summoning}, right). We implemented several example applications, including a photo browser, music player, and a social media feed, to illustrate the breadth of possibilities here.

\subsection{Pointing and Scrolling}
\begin{figure}[h]
  \centering
  \includegraphics[width=\linewidth]{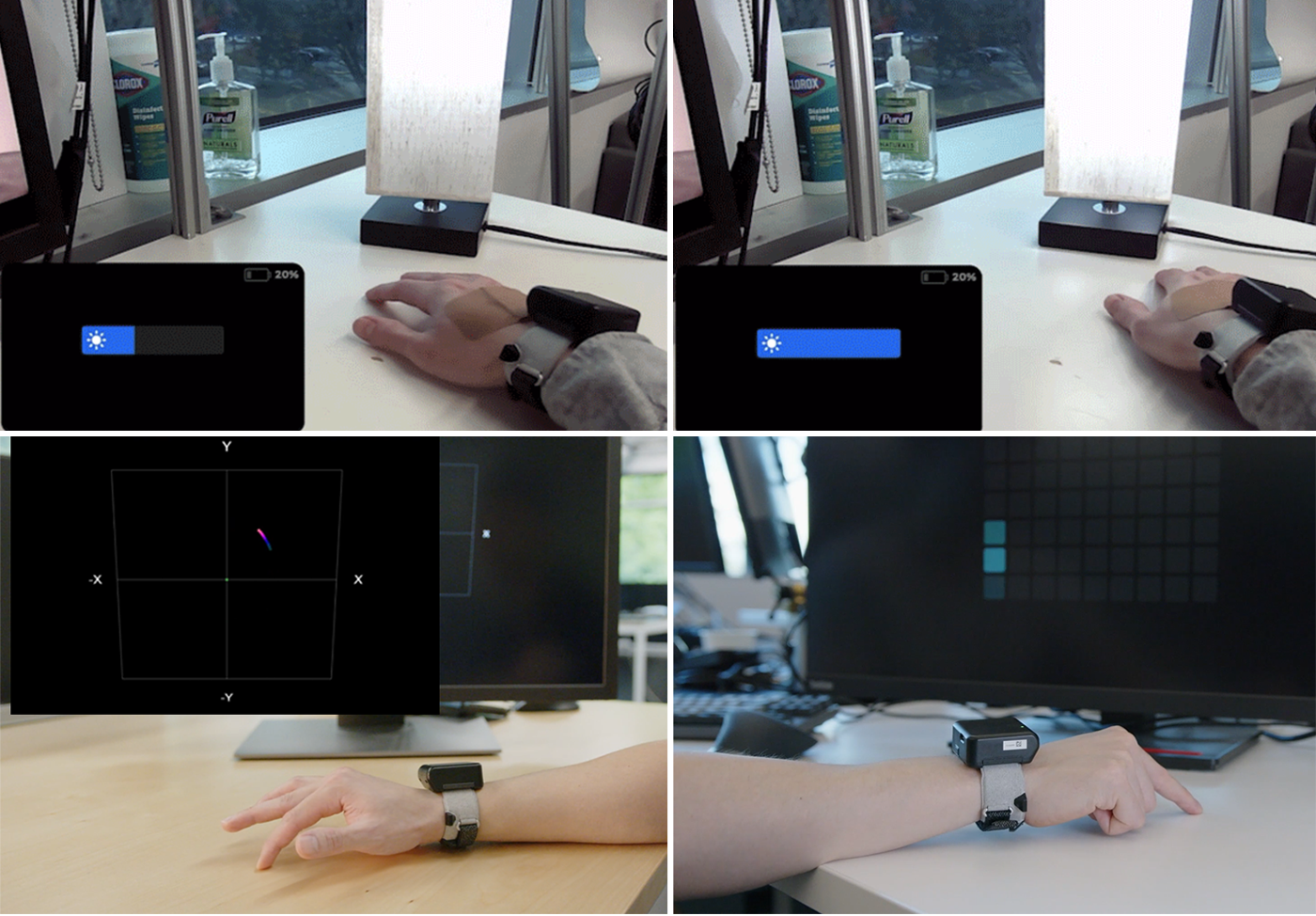}
  \caption{(top) A user controls the lamp brightness by sliding along a surface, highlighting the ability of \sysname{} to enable 1D continuous input. Continuous (bottom, left) and discrete (bottom, right) forms of 2D pathing applications that are enabled by \sysname{}.}
  \label{fig:pointing}
  \Description{}
\end{figure}

Although basic discrete tap and directional gestures are sufficient for many applications, we can enable higher-bandwidth interaction using continuous input. In this exploration, we consider how both 1D and 2D continuous input can enable more fluid interactions. In our first example, shown in Figure~\ref{fig:pointing}~(top), we demonstrate an IoT application where a user can slide their hand left and right next to a smart light to adjust the brightness. With a quick roll of the wrist, they can switch between adjusting the brightness and hue of the light. A tap on the surface confirms the adjustment. By restricting input to one dimension, this interaction provides a simple, reliable control scheme.

\begin{figure*}[h]
  \centering
  \includegraphics[width=6in]{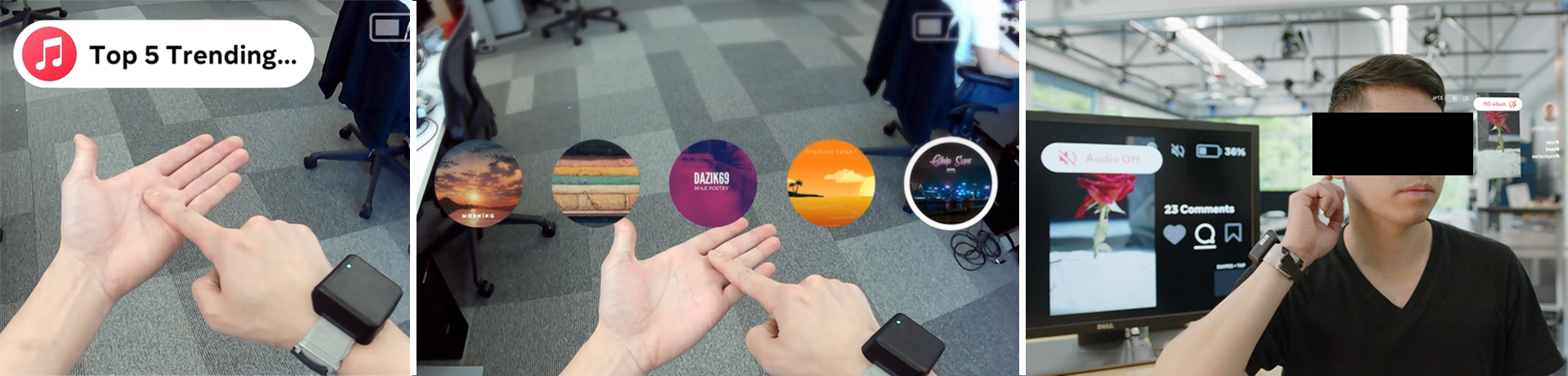}
  \caption{\sysname{} enables on-body controls for always-available interactions, and we can use body pose estimates to adapt the functionality of touches based on their location. In this application, a user can tap on the palm to respond to a notification and drag along the palm to select from a list of suggested songs. Mute and unmute functionality are triggered by tapping behind the ear.}
  \label{fig:int_onbody}
  \Description{}
\end{figure*}

Our second example, shown in Figure~\ref{fig:pointing}~(bottom), tests the limits of \sysname{}'s capabilities by enabling 2D trackpad input. In this example, we use the fingertip tracking fused with inertial forearm tracking to derive a 2D cursor estimate. We show how a continuous estimate can be used to allow a user to path over a 2D grid. 2D pathing from a wristband is a highly challenging task and though \sysname{} is not yet comparable to commercial laptop trackpads, which are optimized for precision pointing tasks, we found that it is possible to draw coarse shapes or select from a small (e.g., 10x10) grid.

\subsection{Contextual touch}
We also consider how the location context of touch interactions can enable dynamic, responsive interfaces. 
By combining \sysname{}'s ability to estimate upper body pose with its ability to distinguish touch on the body from touch on environmental surfaces, we can design touch experiences that adapt based on where you touch. 

In our first example, shown in Figure~\ref{fig:int_onbody}, we extend the universal control example to place a music player interface on the palm. Using the bioimpedance module, users can summon the touch controller on their opposing palm. This offers an always-available touch interface, even if there is no flat surface nearby. Users can then scroll through a one dimensional music playlist by moving the edge of their finger across their opposing palm. Haptics on the device provide a sensation of detents. A one second dwell time threshold is used to enable song selection. To leverage body context for input, we also add a mute/unmute shortcut when the user taps behind their ear or on the side of their face.

\begin{figure}[h]
  \centering
  \includegraphics[width=2in]{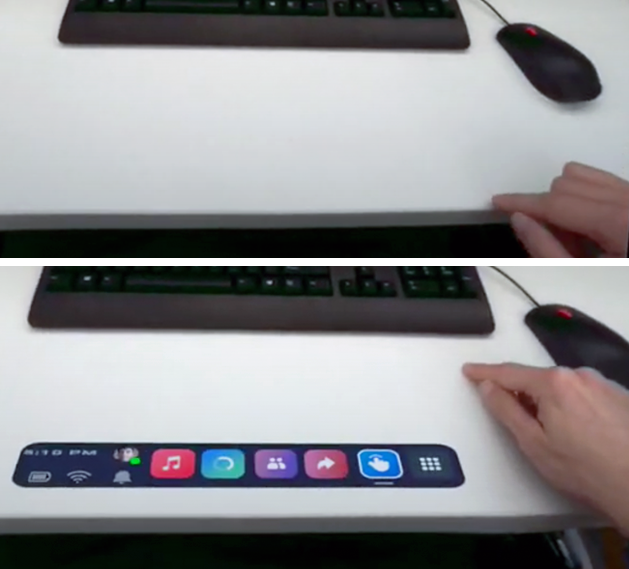}
  \caption{The optical ToF sensor on \sysname{} can detect pose relative to surface edges. By swiping up along the edge of the table, a user can summon an anchored touch interface.}
  \label{fig:int_edge}
  \Description{}
\end{figure}

In our second example, shown in Figure~\ref{fig:int_edge}, we showcase how we can activate contextually relevant interfaces as we approach a surface. By leveraging \sysname{}’s optical sensors, we can detect the proximity or edge of a surface to anchor relevant interfaces. For instance, a user can swipe from the edge towards the center of a table to activate a horizontally anchored interface, much like we activate interfaces on our touchscreen devices.

\subsection{In-Air and Touch Transitions}
While \sysname{} focuses on leveraging environmental surfaces and our body for touch input, we are not limited to performing interactions on surfaces. For instance, when a user is on the go, they may benefit from a wider range of controls and semantic mappings on or around their body in addition to an ad-hoc surface for input.

\begin{figure}[h]
  \centering
  \includegraphics[width=\linewidth]{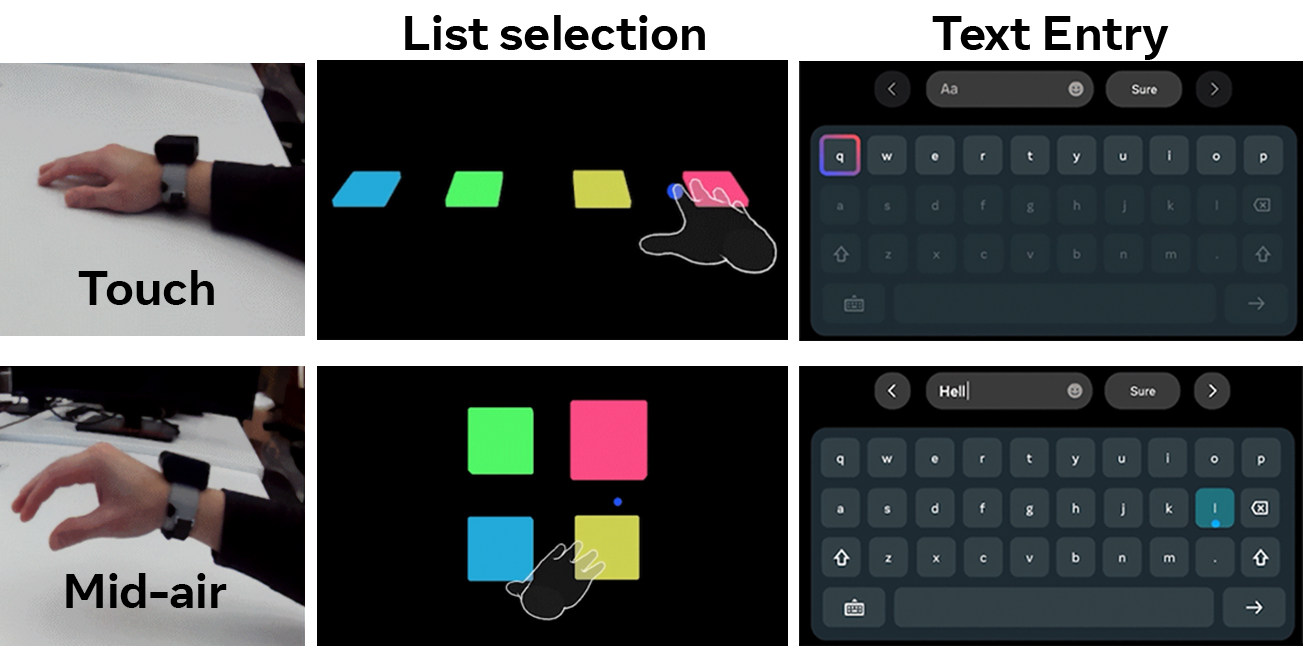}
  \caption{Interfaces can be made responsive based on whether the hand is touching a surface. (middle) Here we transition automatically between a linear 4-item list for easy left-right selection while the hand is on a surface, and a 2x2 grid for mid-air pointing when no surface is detected. (right) A responsive interface for a text entry task. Control adapts from a touch-based keyboard and a mid-air pointing-based keyboard.}
  \label{fig:int_transitions}
  \Description{}
\end{figure}

In our first example, we demonstrate how interfaces can respond based on hand pose and touch state. When the hand is resting on the surface, we present a 4-item selection interface~(Figure~\ref{fig:int_transitions}, left). This responsively becomes a 2x2 grid that can be controlled by pointing / wrist deflection and pinch when the user’s hand is in mid-air~(Figure~\ref{fig:int_transitions}).

In our second example, we demonstrate multiple methods a user can leverage to interact with a virtual keyboard. Users can change the control mechanics by simply moving between surface and in-air modes for input. For instance, placing a user’s hand on a surface for two seconds enables keyboard mode. Users can swipe to select a row and use the inertial body scale touch tracking for character selection within that row. Alternatively, users can perform a selection in air using wrist deflection and dwell.

\section{Discussion} \label{sec:discussion}

Designed as a platform to explore a wide research space, these applications offer glimpses of a rich, contextually-adaptive interaction language that users found to be highly learnable, with minimal cognitive/physical demand. This supports our hypothesis that wristband-driven sensing can deliver always-available touch capabilities in a low-friction manner. 

We adopted a multimodal approach, with each sensor carefully selected to capture a unique aspect of touch interactions.
Because \sysname{} distinguishes between a variety of touch contexts, we can enable a much more intuitive interaction language beyond a typical smartwatch touchscreen or tap and swipe microgestures; for example, in Section~\ref{sec:interactions} we show how tapping the ear can mute audio, the palm can used to navigate a list, and content can be embedded in the environment near relevant objects or on surface edges.

The limitations of abstract microgestures or touch gestures on small smartwatch screens or temple arms of glasses become apparent when trying to create a more expressive interaction language. The limited surface area limits the gestures that can be performed, leading to a reliance on abstract menu navigation or complex gestural languages. In contrast, by situating interactions in the environment or on the body, we increase expressiveness and provide a more intuitive way to interact with digital information.

Notably, \sysname{} is the first wristband to provide a complete set of touch primitives and exhibits state-of-the-art single-wristband performance on continuous touch state and tracking tasks.
Overall, we found that our models relied heavily on proximity and inertial information - the proximity sensors provide strong motion cues for touch/release and wrist deflection and the IMU provides a strong indicator of impact and motion.
Moreover, these signals are simple and highly correlated with the actions of interest.
The bioimpedance and sEMG sensors provide specialized signals, particularly for body touch estimation. However, we expect their benefits will be more apparent when fused and scaled to more adversarial in-the-wild conditions.

Though the current model performance is promising, there is often a sizeable gap between performance demonstrated in HCI literature and performance requirements for reliable use across a broad population in new environments.
This gap can be attributed, in part, to the substantial variability in data captured across a large population, encompassing physiological differences such as wrist size and shape, skin color and conductivity, hair distribution, and individual nuances in device wear and gesture performance.
Learned models for novel wearable sensing systems are often trained on relatively small datasets with less than 25 participants~\cite{kim2023vibaware,meier2021tapid,streli2022taptype,shen2024mousering,gong2020acustico,waghmare2023zring,electroring}; however, recent work has found that gestural models in such contexts continue to improve in performance even beyond thousands of participants’ worth of data~\cite{ctrl2024generic,hayashi2021radarnet}.
In particular, sEMG has been shown to require thousands of participants to achieve generalized performance on gesture tasks~\cite{ctrl2024generic}.
This gives us confidence that the performance we're already seeing on our combined 100 participant dataset will continue to improve as data collection is scaled.
Beyond increasing the size of the dataset, we further expect that personalized (e.g. fine-tuned) models can adapt to the wearer and the specific features of their input.

Though some of the variability can be captured by improving data collections and models, for some users, performance could be improved simply by improving wristband fit.
Variability in wrist anatomy makes some electrodes subject to liftoff, significantly impacting model performance.
We found that the body touch model was particularly susceptible to variable electrode contact and currently requires a snug fit for proper performance.
We also developed two sizes of the wristband to minimize variability in sensor placement on different sized wrists, but as a research prototype, \sysname{} is still somewhat bulky and we expect that if made smaller and more conformable, we would see fewer contact-related issues.

In addition to these design considerations, there are also future system-related opportunities to make \sysname{} more usable and reliable. 
For example, the battery life on our current prototype is only optimized for data collection sessions and demos on the order of 1 hour long. While we showcased the ability to sense nearby surface contact to toggle costlier subsystems, we further expect that limiting sensing based on application context will limit the relative on-time of most sensors. There are also clear opportunities to improve the sensing architectures to reduce power; for example, the bioimpedance module (see Table~\ref{tab:sensors}) power could be reduced significantly by exploring alternative demodulation approaches.
\sysname{} also operates by streaming all data over Bluetooth with all processing on a GPU-enabled desktop PC. Though out of scope for this work, we envision future iterations of these models to prioritize fitting within the compute envelope of a typical smartwatch.

In this work, we explored a standalone sensing model, with input driven from \sysname{}’s capabilities in isolation. However, we expect the real power of \sysname{} to be unlocked when working within a constellation of devices, such as smart glasses or an XR headset. Specifically, \sysname{} excels at solving the last-millimeter problem. As such, it effectively complements egocentric vision-based sensing approaches, e.g. from an HMD. Taking touch tracking as an example, a vision-based system could capture coarse motion reliably and without drift (e.g. correcting for most of the error observed via the wrist tracking model), while \sysname{} could provide finer details of finger/wrist deflection and precise touch state. The fusion of these signals into a single, high-quality estimate of touch remains exciting future work. 

As additional future work, we aim to further explore the capabilities of \sysname{} in mobile contexts. To make this problem tractable with moderate sized datasets, we focused this work on stationary contexts (e.g., sitting or standing at a table), but we believe there is significant value for touch interactions on the go, where you can co-opt any nearby surface for input. 
In these less-constrained contexts, it is important to mitigate false positives when performing everyday tasks, like grasping objects.
We anticipate a wake model, much like the double tap to summon functionality described in Section~\ref{sec:interactions}, or application-driven wake to be critical for usability.

\section{Conclusion} \label{sec:conclusion}
In this work, we introduce \sysname{}, a research platform that demonstrates how multimodal sensing from the wrist can enable contextually-adaptive touch interactions anywhere. \sysname{} combines optical, inertial, and sEMG modalities alongside a novel bioimpedance sensor to provide touch state and touch tracking capabilities. We then show how these primitives combine to enable a rich language of interactions on surfaces and on the body. Though work remains to generalize this performance across a broad population and in mobile contexts, we look forward to a future where an always-on wristband can complement a constellation of other wearable AR/AI devices to support context-aware interaction with the physical world.

\begin{acks}
We thank the entire extended team for their contributions to this work. In particular, we thank Bao Tung Tran and Dan Porter for their engineering contributions and leadership, Brennon Costello for support on the bioimpedance module, Shunbao Chen and Casey Brown for supporting data collections, Austin Lee and James Tichenor for inspired design contributions, and Neil Weiss for protocol design and evaluation support.
\end{acks}

\bibliographystyle{ACM-Reference-Format}
\bibliography{0-bibliography}
\appendix
\onecolumn
\clearpage
\section{Electrical System Architecture} \label{sec:appendix-sys}

\noindent\begin{minipage}{\textwidth}
    \centering
    \includegraphics[width=6in]{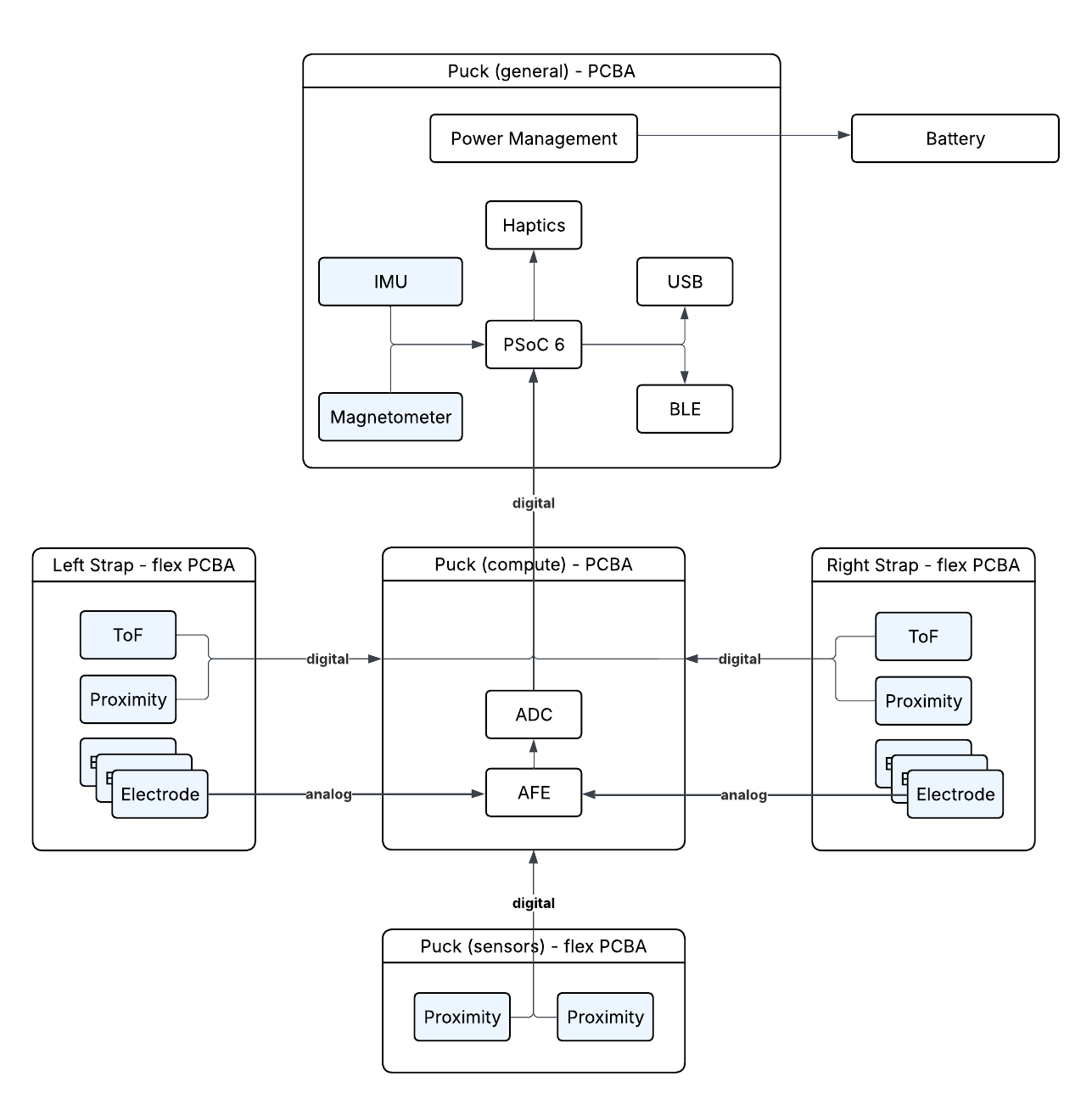}
    \captionof{figure}{Additional details on \sysname{} electrical system architecture}
    \label{fig:appendix-arch}
\end{minipage}

\clearpage
\section{Additional Model Details} \label{sec:appendix-models}

All of the learned models follow a similar CNN + LSTM architecture to capture temporal patterns across sensor channels. Unless otherwise noted, no featurization was performed and models operated on raw signals. Because our system operates on fairly long, overlapping sliding windows of sensor data, the LSTM hidden state is not preserved between windows. We expect that adapting this architecture to preserve hidden state will improve both computational efficiency and model performance, but we leave this to future work. Each model has small architectural difference, which we describe and depict below.

\subsection{On-World Touch State Model}

As shown in Figure~\ref{fig:nn-world}, the on-world touch state model consists of three layers of 1D convolution interspersed with max-pooling layers. Convolution filter sizes were 67, 23, 7 with counts of 32, 32, and 16, and strides of 5, 3, and 1, respectively. The two 1D max-pooling layers had a size of 3 and 2, respectively. We used an LSTM with a hidden state of size 128, followed by a fully connected layer with softmax activation.
\\
\noindent\begin{minipage}{\textwidth}
\centering
\begin{tikzpicture}
\tikzstyle{connection}=[ultra thick,every node/.style={sloped,allow upside down},draw=\edgecolor,opacity=0.7]
\tikzstyle{copyconnection}=[ultra thick,every node/.style={sloped,allow upside down},draw={rgb:blue,4;red,1;green,1;black,3},opacity=0.7]

\pic[shift={(0,0,0)}] at (0,0,0)
    {Box={
        name=window,
        caption=Input,
        xlabel={{18, }},
        zlabel=2000,
        fill=\UnpoolColor,
        height=2,
        width=2,
        depth=40
        }
    };

\pic[shift={(1,0,0)}] at (window-east)
    {Box={
        name=conv1,
        caption=Conv1D (18x67),
        xlabel={{32, }},
        zlabel=387,
        fill=\ConvColor,
        height=2,
        width=4,
        depth=20
        }
    };

\pic[shift={ (1,0,0) }] at (conv1-east)
    {Box={
        name=pool1,
        caption=Max\\Pool,
        fill=\PoolColor,
        opacity=0.5,
        height=2,
        width=4,
        depth=13
        }
    };

\pic[shift={(0,0,0)}] at (pool1-east)
    {Box={
        name=conv2,
        caption=Conv1D\\(32x23),
        xlabel={{32, }},
        zlabel=36,
        fill=\ConvColor,
        height=2,
        width=4,
        depth=4
        }
    };

\pic[shift={ (1,0,0) }] at (conv2-east)
    {Box={
        name=pool2,
        caption=Max\\Pool,
        fill=\PoolColor,
        opacity=0.5,
        height=2,
        width=4,
        depth=6
        }
    };

\pic[shift={(0,0,0)}] at (pool2-east)
    {Box={
        name=conv3,
        caption=Conv1D\\(32x7),
        xlabel={{32, }},
        zlabel=12,
        fill=\ConvColor,
        height=2,
        width=4,
        depth=3
        }
    };

\pic[shift={(1,0,0)}] at (conv3-east)
    {Box={
        name=lstm,
        caption=LSTM,
        xlabel={{128, }},
        zlabel=,
        fill=\LstmColor,
        height=2,
        width=16,
        depth=1
        }
    };

\pic[shift={(1,0,0)}] at (lstm-east)
    {Box={
        name=fc1,
        caption=FC,
        xlabel={{ "3", "dummy" }},
        zlabel=,
        fill=\FcColor,
        opacity=0.8,
        height=2,
        width=1,
        depth=1
        }
    };

\end{tikzpicture}

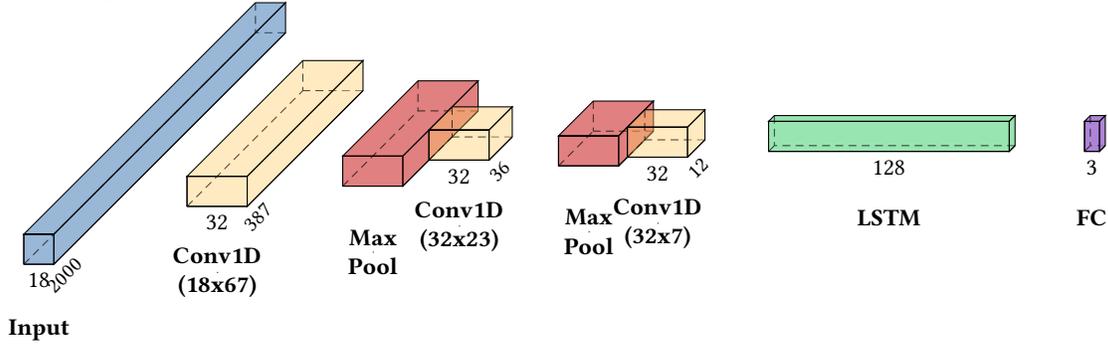
\captionof{figure}{Network architecture for on-world touch state model}
\label{fig:nn-world}
\end{minipage}

\subsection{On-Body Touch State Model}

As shown in Figure~\ref{fig:nn-body}, the on-body touch state model is nearly identical to the on-world model, but with an initial 2x downsampling layer and with filter sizes of 35, 15, and 9, counts of 60, 60, and 30, and strides of 1. The two 1D max-pooling layers each had a size of 2. Unlike before, we use the LSTM output (hidden state size of 60) from each time step as input to a fully connected layer with log softmax activation.

\noindent\begin{minipage}{\textwidth}
\centering
\begin{tikzpicture}
\tikzstyle{connection}=[ultra thick,every node/.style={sloped,allow upside down},draw=\edgecolor,opacity=0.7]
\tikzstyle{copyconnection}=[ultra thick,every node/.style={sloped,allow upside down},draw={rgb:blue,4;red,1;green,1;black,3},opacity=0.7]

\pic[shift={(0,0,0)}] at (0,0,0)
    {Box={
        name=window,
        caption=Input,
        xlabel={{4, }},
        zlabel=500,
        fill=\UnpoolColor,
        height=2,
        width=1,
        depth=40
        }
    };

\pic[shift={(.5,0,0)}] at (window-east)
    {Box={
        name=window2,
        caption=Down\\sample,
        xlabel={{4, }},
        zlabel=250,
        fill=\UnpoolColor,
        height=2,
        width=1,
        depth=20
        }
    };

\pic[shift={(1,0,0)}] at (window2-east)
    {Box={
        name=conv1,
        caption=Conv1D\\(18x67),
        xlabel={{60, }},
        zlabel=216,
        fill=\ConvColor,
        height=2,
        width=6,
        depth=17
        }
    };

\pic[shift={ (1,0,0) }] at (conv1-east)
    {Box={
        name=pool1,
        caption=Max\\Pool,
        fill=\PoolColor,
        opacity=0.5,
        height=2,
        width=6,
        depth=11
        }
    };

\pic[shift={(0,0,0)}] at (pool1-east)
    {Box={
        name=conv2,
        caption=Conv1D\\(32x23),
        xlabel={{60, }},
        zlabel=94,
        fill=\ConvColor,
        height=2,
        width=6,
        depth=9
        }
    };

\pic[shift={ (1,0,0) }] at (conv2-east)
    {Box={
        name=pool2,
        caption=Max\\Pool,
        fill=\PoolColor,
        opacity=0.5,
        height=2,
        width=6,
        depth=7
        }
    };

\pic[shift={(0,0,0)}] at (pool2-east)
    {Box={
        name=conv3,
        caption=Conv1D\\(32x7),
        xlabel={{30, }},
        zlabel=39,
        fill=\ConvColor,
        height=2,
        width=3,
        depth=5
        }
    };

\pic[shift={(.5,0,0)}] at (conv3-east)
    {Box={
        name=lstm,
        caption=LSTM,
        xlabel={{60, }},
        zlabel=,
        fill=\LstmColor,
        height=2,
        width=6,
        depth=1
        }
    };

\pic[shift={(.5,0,0)}] at (lstm-east)
    {Box={
        name=fc1,
        caption=FC,
        xlabel={{ "39", "dummy" }},
        zlabel=60,
        fill=\FcColor,
        opacity=0.8,
        height=2,
        width=3,
        depth=1
        }
    };

\pic[shift={(.5,0,0)}] at (fc1-east)
    {Box={
        name=fc2,
        caption=FC,
        xlabel={{ "3", "dummy" }},
        zlabel=1,
        fill=\FcColor,
        opacity=0.8,
        height=2,
        width=1,
        depth=1
        }
    };

\end{tikzpicture}


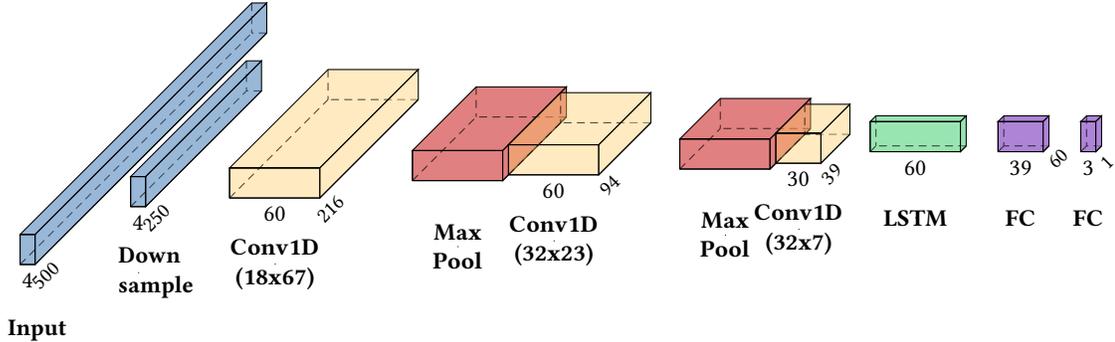
\captionof{figure}{Network architecture for on-body touch state model}
\label{fig:nn-body}
\end{minipage}

\subsection{Fingertip Touch Tracking Model}
The fingertip touch tracking model, shown in Figure~\ref{fig:nn-trackpad}, treats sEMG data and IR/IMU data separately. EMG data is featured using covariance features. IMU+IR data is downsampled by a factor of 2 and passed through two 1D depthwise separable convolution layers with filter sizes of 11 and 5, respectively, a stride of 1, and counts of 40. Each of these is followed by a max pooling layer with a size of 2. The outputs of both paths are flattened and passed through two fully connected layers of size 20 and 2 with ReLU activation.

\noindent\begin{minipage}{\textwidth}
\centering

\begin{tikzpicture}
\tikzstyle{connection}=[ultra thick,every node/.style={sloped,allow upside down},draw=\edgecolor,opacity=0.7]
\tikzstyle{copyconnection}=[ultra thick,every node/.style={sloped,allow upside down},draw={rgb:blue,4;red,1;green,1;black,3},opacity=0.7]

\pic[shift={(0,3,0)}] at (0,0,0)
    {Box={
        name=emg,
        caption=sEMG,
        xlabel={{8, }},
        zlabel=400,
        fill=\UnpoolColor,
        height=2,
        width=2,
        depth=20
        }
    };

\pic[shift={(.5,0,0)}] at (emg-east)
    {Box={
        name=feature,
        caption=Feature,
        xlabel={{64, }},
        zlabel=31,
        fill=\UnpoolColor,
        height=2,
        width=4,
        depth=8
        }
    };

\pic[shift={(0,0,0)}] at (0,0,0)
    {Box={
        name=imu,
        caption=IR + IMU,
        xlabel={{10, }},
        zlabel=400,
        fill=\UnpoolColor,
        height=2,
        width=2,
        depth=20
        }
    };

\pic[shift={(.5,0,0)}] at (imu-east)
    {Box={
        name=down,
        caption=Down\\sample,
        xlabel={{10, }},
        zlabel=200,
        fill=\UnpoolColor,
        height=2,
        width=2,
        depth=10
        }
    };

\pic[shift={(.5,0,0)}] at (down-east)
    {Box={
        name=conv1,
        caption=Conv1D\\(40x90),
        xlabel={{40, }},
        zlabel=90,
        fill=\ConvColor,
        height=2,
        width=5,
        depth=5
        }
    };

\pic[shift={ (.5,0,0) }] at (conv1-east)
    {Box={
        name=pool1,
        caption=Max\\Pool,
        fill=\PoolColor,
        opacity=0.5,
        height=2,
        width=5,
        depth=3
        }
    };

\pic[shift={(0,0,0)}] at (pool1-east)
    {Box={
        name=conv2,
        caption=Conv1D\\(40x41),
        xlabel={{40, }},
        zlabel=41,
        fill=\ConvColor,
        height=2,
        width=5,
        depth=3
        }
    };

\pic[shift={ (.5,0,0) }] at (conv2-east)
    {Box={
        name=pool2,
        caption=Max\\Pool,
        fill=\PoolColor,
        opacity=0.5,
        height=2,
        width=5,
        depth=1
        }
    };

\pic[shift={(1,1.5,0)}] at (pool2-east)
    {Box={
        name=flatten1,
        caption=Flatten,
        xlabel={{2784, }},
        zlabel=,
        fill=\UnpoolColor,
        height=20,
        width=1,
        depth=1
        }
    };

\draw [connection]  (pool2-east)    -- node {\midarrow} (flatten1-west);

\draw [connection]  (feature-east)    -- node {\midarrow} (flatten1-west);

\pic[shift={(.5,0,0)}] at (flatten1-east)
    {Box={
        name=fc1,
        caption=FC,
        xlabel={{ "20", "dummy" }},
        zlabel=,
        fill=\FcColor,
        opacity=0.8,
        height=3,
        width=1,
        depth=1
        }
    };

\pic[shift={(.5,0,0)}] at (fc1-east)
    {Box={
        name=fc2,
        caption=FC,
        xlabel={{ "2", "dummy" }},
        zlabel=,
        fill=\FcColor,
        opacity=0.8,
        height=1,
        width=1,
        depth=1
        }
    };

\end{tikzpicture}
\captionof{figure}{Network architecture for trackpad model}
\label{fig:nn-trackpad}
\end{minipage}
\clearpage
\section{Additional Evaluation Details and Results} \label{sec:appendix-study}

\begin{figure*}[h]
  \centering
  \includegraphics[width=4in]{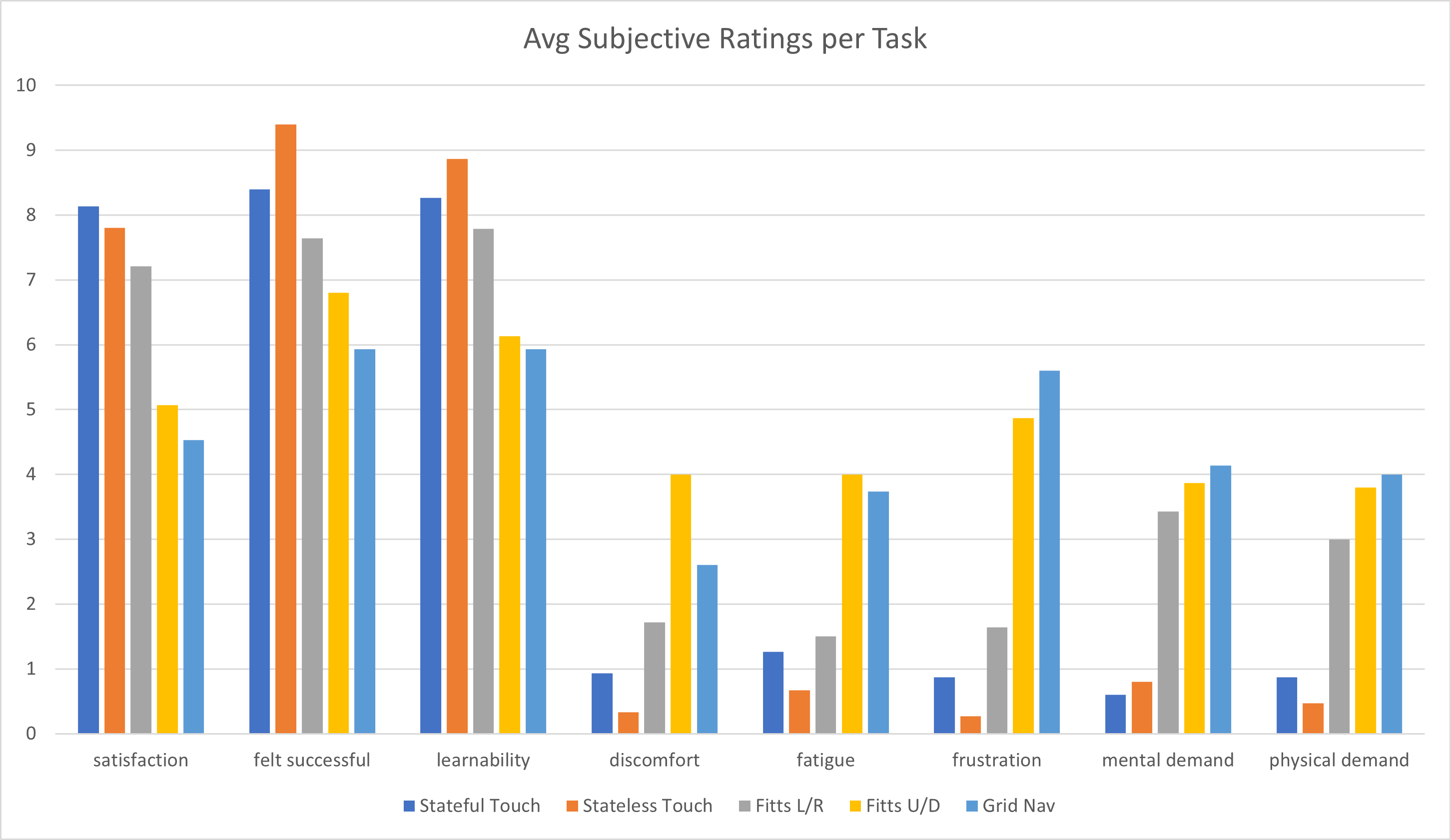}
  \caption{Reported subjective measures by gesture/task type.}
  \label{fig:subjective-ratings}
  \Description{}
\end{figure*}

\begin{figure*}[h]
  \centering
  \includegraphics[width=6in]{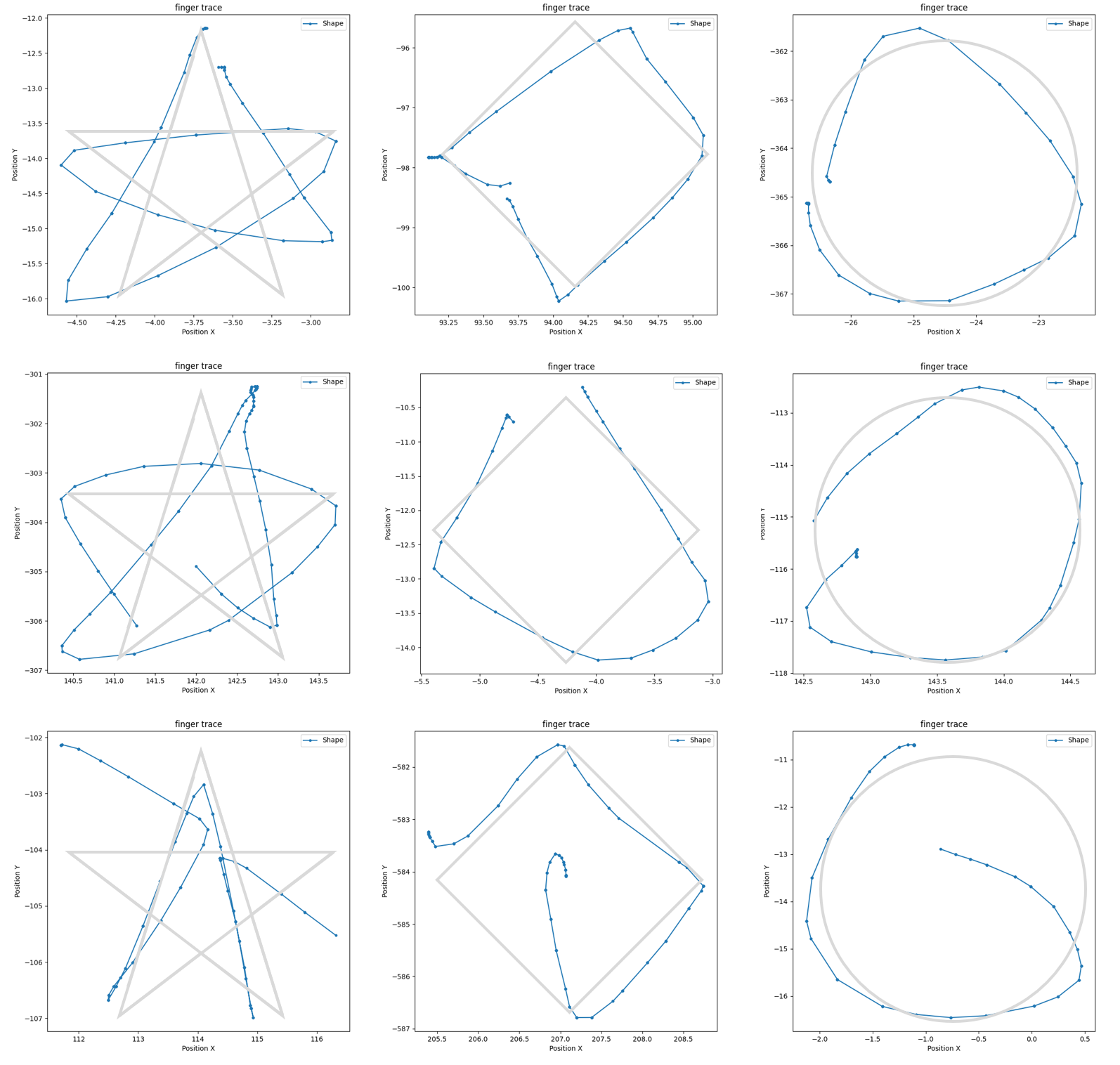}
  \label{fig:tracking-traces-all}
  \caption{Additional examples of traces reconstructed with the 2D trackpad model.}
  \Description{}
\end{figure*}

\begin{figure*}[h]
  \centering
  \includegraphics[width=6in]{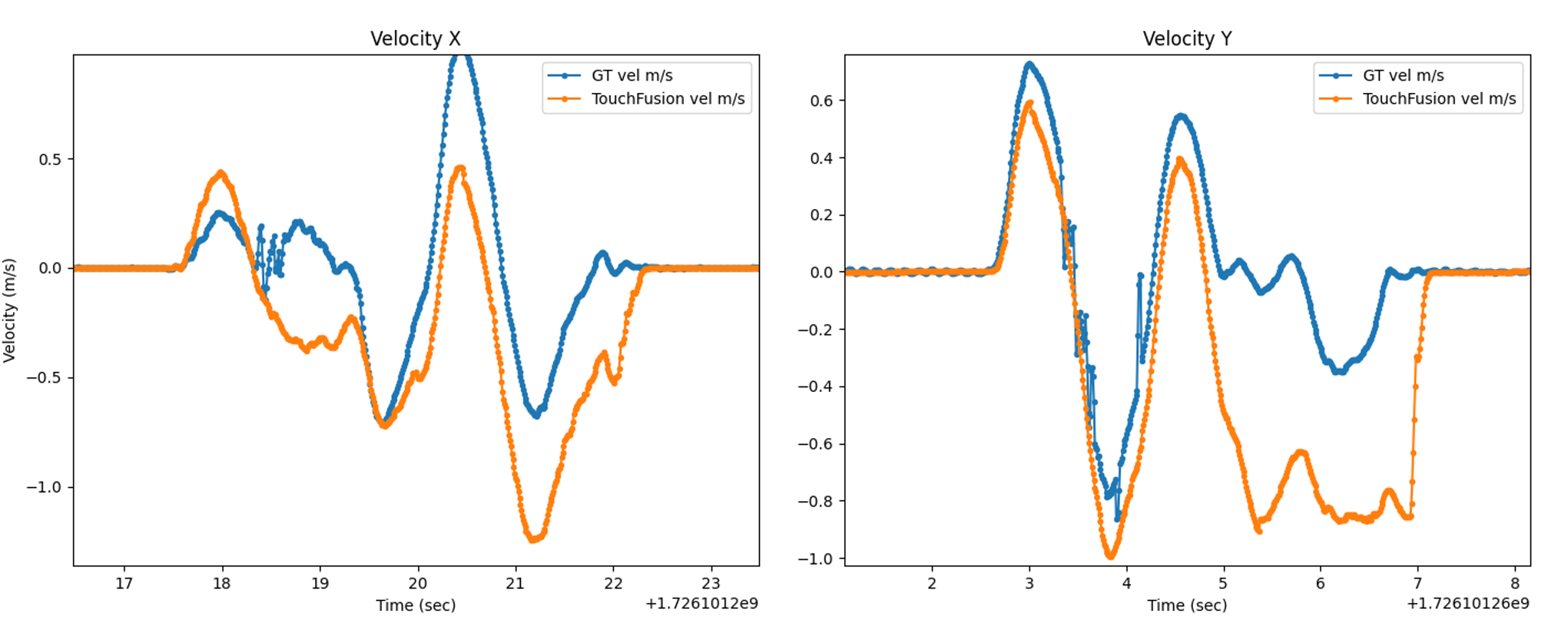}
  \label{fig:tracking-p7}
  \caption{Examples of worst-case tracking failures for the dead-reckoning wrist tracking model.}
  \Description{}
\end{figure*}

\begin{figure*}[h]
  \centering
  \includegraphics[width=6in]{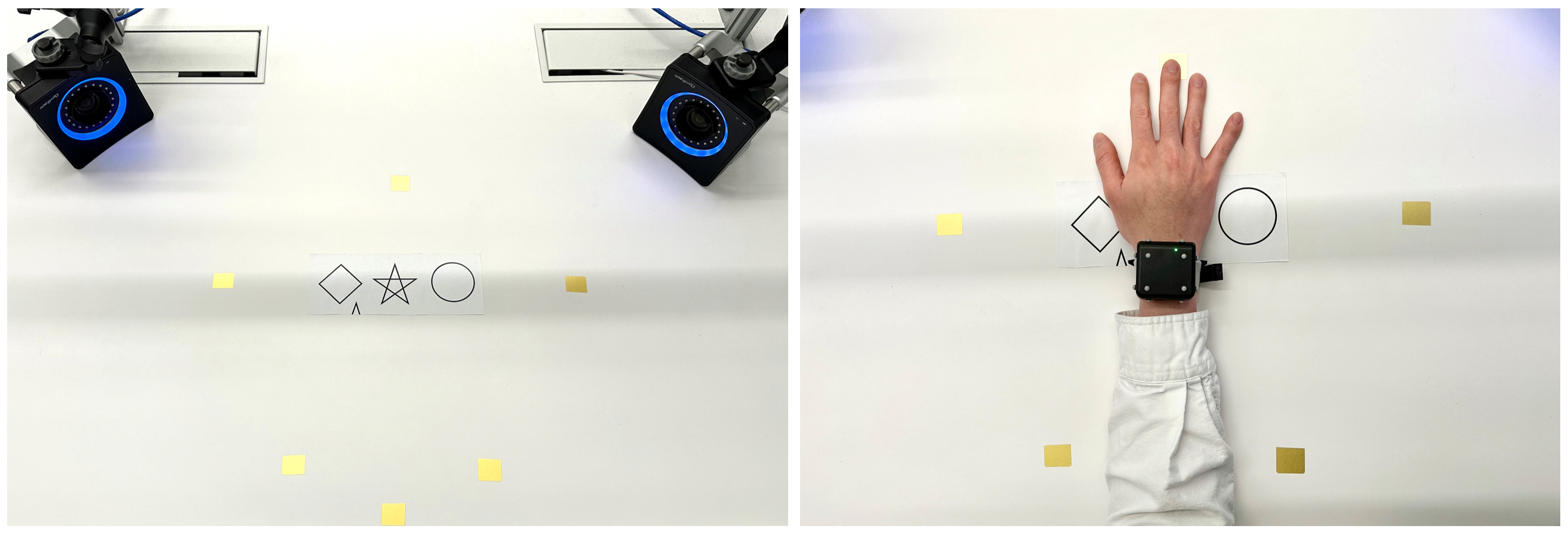}
  \caption{Study setup for wrist + fingertip tracking evaluation}
  \Description{}
  \label{fig:eval}
\end{figure*}


\end{document}